\documentclass{jpp}
\usepackage{graphicx}
\usepackage{ulem}
\usepackage{float}
\usepackage[utf8]{inputenc}
\usepackage[T1]{fontenc}
\usepackage{xcolor}
\usepackage{tikz}
\DeclareRobustCommand{\dashedline}{\tikz[baseline=-0.6ex]\draw[dashed, ultra thick] (0,0)--(0.6,0);}
\DeclareRobustCommand{\dottedline}{\tikz[baseline=-0.6ex]\draw[dotted,  thick] (0,0)--(0.45,0);}

\definecolor{iotaHigh}{HTML}{1B9E77}
\definecolor{iotaMedium}{HTML}{7570B3}
\definecolor{iotaLow}{HTML}{D95F02}
\definecolor{starblue}{HTML}{1F77B4}
\definecolor{starred}{HTML}{D62728}

\newcommand{\colorrect}[1]{%
    \tikz[baseline=-0.6ex]{%
        \fill[#1, rounded corners=2pt] (0,-3pt) rectangle (12pt, 4pt);%
    }%
}

\usepackage{macros}
\usepackage{orcidlink}
\usepackage{siunitx}

\DeclareMathSymbol{\sm}{\mathbin}{AMSa}{"39}

\usepackage{enumitem} %
\setlist[enumerate]{topsep=\halflineskip,itemsep=0pt,partopsep=1ex,parsep=0pt}

\shorttitle{A first STAR\_Lite coil configuration}
\shortauthor{G. Harrer, A. Giuliani, M. Padidar, R. Davies, S. Naik, C. Lowe}

\title{STAR\_Lite: A stellarator designed to experimentally validate non-resonant divertors}

\author{Georg Harrer\orcidlink{0000-0002-1150-3987}\aff{1}, 
  \corresp{\email{fusion@hamptonu.edu}},
  Andrew Giuliani\orcidlink{0000-0002-4388-2782}\aff{2}, Misha Padidar\orcidlink{0000-0002-0710-4377}\aff{2}, Robert Davies\orcidlink{0000-0001-5570-5882}\aff{3}, Shibabrat Naik\orcidlink{0000-0001-7964-2513}\aff{4}, \and Calvin Lowe\aff{1}}

\affiliation{\aff{1}Department of Physics, Hampton University, Hampton, VA, 23663, USA
\aff{2}Center for Computational Mathematics, Flatiron Institute, New York, NY, 10010, USA
\aff{3}Max Planck Institute for Plasma Physics, Wendelsteinstraße 1, 17491 Greifswald, Germany
\aff{4}Department of Mathematics, Hampton University, Hampton, VA 23663, USA}

\begin{document}
\maketitle
\begin{abstract}
The non-resonant divertor (NRD) offers a promising exhaust solution for stellarators, combining topological simplicity with resilience to magnetic field perturbations. To experimentally validate the robustness of non-resonant divertors in a quasi-axisymmetric (QA) configuration, we introduce STAR\_Lite, a new stellarator experiment at Hampton University. This paper details the design and analysis of the first STAR\_Lite coil configuration, STAR\_Lite-A. 
The two field-period configuration manifests an NRD through X-points with zero rotational transform, at the top and bottom of the device. The divertor legs extruding from the X-points are topologically similar to the poloidal divertors of tokamaks. 
To expand the experimental range, STAR\_Lite-A is optimized for experimental flexibility, producing a wide range of distinct QA configurations by only varying the currents in the modular coils.
The NRDs not only persist across these configurations, but numerical strike-line simulations confirm that heat exhaust remains resilient to changes in coil currents, with plasma following the divertor legs and creating a toroidal, discontinuous, strike pattern. 
We further examine the resilience of the NRD to magnetic perturbations caused by manufacturing errors in the modular coils. 
We find that quasisymmetry and the existence of X-points is well-preserved under these magnetic field changes, but the rotational transform may vary substantially and displacements of the divertor X-points may lead to one X-point having a dominant effect on edge transport. Overall, our analysis indicates a compact, modular design can likely generate a resilient NRD structure while satisfying the practical constraints of a university-scale experiment.
\end{abstract}

\section{Introduction}
\label{sec:introduction}
The stellarator fusion concept is experiencing a renaissance fueled by the success of Wendelstein 7-X (W7-X) \citep{Wolf_2017, Klinger_2019}, and enabled by theoretical and computational advances in obtaining quasi-symmetric and quasi-isodynamic configurations under various engineering constraints ~\citep{gates2017,landreman_simsopt_2021,landreman_magnetic_2022,wechsung_precise_2022,goodman_constructing_2022}. %
An essential requirement of a commercially viable stellarator reactor is a divertor, which handles the escaping plasma, exhausting heat and steadily removing helium ``ash'' and impurities \citep{burnett1958divertor, konig2002divertor, gates2018stellarator}.
One divertor concept is the island, or ``resonant'' divertor, such as used by Wendelstein 7-X (W7-X) \cite{renner2004physical}, which relies on a chain of magnetic islands at the plasma edge to divert escaping plasma away from the confined region. The experimental validation of the resonant island divertor in W7-X has established island divertors as a leading reactor-scale divertor concept~\citep{Warmer_2022, Proxima_Stellaris, hegna_infinity_2025, goodman_squid_2025}.

However, recent W7-X experiments have highlighted challenges with the island divertor, specifically regarding neutral particle compression and a complex dependence of pumping efficiency on island geometry \citep{Kremeyer_2022}. Additionally, the resonant divertor poses operational challenges: it relies on precise control of the edge rotational transform, which may be difficult to maintain as plasma currents evolve \citep{gates2018stellarator, gao2019effects}. %

Alternatives to the island divertor include the helical divertor in the Large Helical Device (LHD)~\citep{Ohyabu_1994} or the ``non-resonant divertor'' (NRD), a label which has been applied to the Helically Symmetric Experiment (HSX)~\citep{Bader_2017}, the Compact Toroidal Hybrid (CTH) \citep{bader2018minimum, Garcia2023CTH}, the National Compact Stellarator Experiment
(NCSX) and the ARIES-CS reactor concept \citep{boozer_2015}, the WISTELL-A stellarator design \citep{bader_advancing_2020}, some W7-X configurations \citep{boeyaert2025analysis} and to a class of stellarator Hamiltonian models \citep{Boozer_2018, Punjabi_2020, punjabi2022magnetic}. NRDs are often characterised as having resilient plasma strike locations (or heat footprints) on the plasma-facing components, which offers an advantage over island divertors. Several studies of different NRD candidates have reported strike locations on the vessel walls falling on a small number of distinct, narrow bands, which are found to be independent of the distance of the wall from the last closed flux surface (LCFS) \citep{punjabi2022magnetic}, coil currents \citep{Bader_2017, Garcia2025HSX}, or plasma current \citep{Garcia2023CTH}. Despite the promising features of NRDs, there exist differences between the defining characteristics of these examples, and there is not a single unified understanding of how this resilient behaviour comes about (briefly reviewed in \Cref{app:nrd_literature}). Consequently, a single precise NRD definition does not exist. In this paper, we adopt the following, functional, definition of a non-resonant divertor: 
\vspace{2pt}
\newline
\textit{An edge magnetic structure which reliably diverts plasma far from the confined region in collimated structures, such that the strike locations on plasma-facing components are a small number of distinct, narrow bands, that are resilient to equilibrium changes.} 
\vspace{2pt}

\begin{figure}
    \centering
    \includegraphics[width=\linewidth]{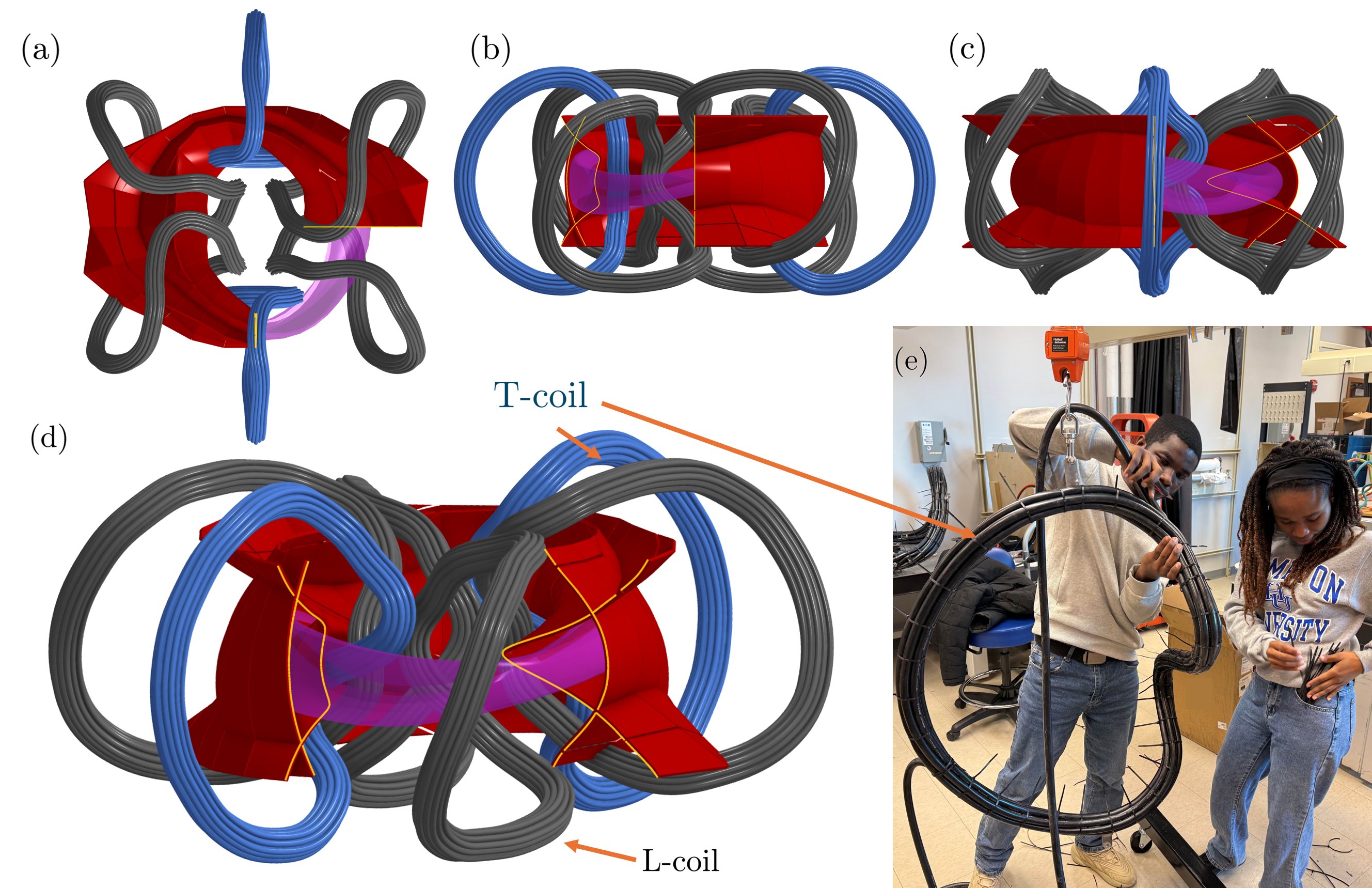}
    \caption{(a-d) \textbf{Design A} viewed from different perspectives; the stable and unstable manifolds emanating from the X-point are shown in red, and the toroidal surface on which quasisymmetry is optimized is shown in purple. The four $L$-coils, and two $T$-coils in black and blue, respectively. In (e), we show students winding a full size $T$-coil at the HU Fusion Lab.}
    \label{fig:designA}
\end{figure}

NRDs following this definition have been studied experimentally in HSX \citep{akerson2016identification} and CTH \citep{allen2023edge, allen2024studies}. 
However, the immense design space of stellarator magnetic topologies remains largely unexplored, suggesting that many NRD-compatible configurations exist beyond those currently reachable in existing devices.
To bridge this gap, we introduce a new, university-scale stellarator experiment, STAR\_Lite (conceptual design in ~\Cref{fig:designA}), 
which is the central experiment of a new fusion research facility at Hampton University that aims to provide a platform for the rapid, cost-effective prototyping and verification of optimized coil geometries \citep{harrer2025star_lite}.
The primary scientific goal of this experiment is to evaluate the resilience of a Non-Resonant Divertor (NRD) within a steady-state quasi-axisymmetric (QA) configuration. To this end, the facility is designed to fulfill three high-level objectives:

\begin{enumerate}[label={(\bfseries Obj. \arabic*)},leftmargin=1.5cm,itemindent=\parindent,itemsep=2pt] 
    \item \, Identify and design a realizable magnetic configuration that possesses the qualities of an NRD: robust diversion of plasma away from the confined region and resilience to changes in plasma equilibrium.
    \label{goal:build}
    \item \, Verify the magnetic structure and divertor features of the built machine satisfy the NRD properties i.e. that an NRD can be deliberately and reliably engineered.
    \label{goal:measurement}

    \item \, Rigorously test the resilience of the NRD structure, i.e. the movement of X-points and strike lines, to various perturbations in a range of magnetic geometries.
    \label{goal:perturb}
\end{enumerate}

This manuscript details the design of the \textit{first} STAR\_Lite coil configuration, \textbf{STAR\_Lite-A}, abbreviated \textbf{design A} (in the future, the facility intends to feature additional coil configurations to fulfill other experimental objectives). \textbf{Design A} is intended to meet the three primary research objectives, while being practically buildable and operable within the means of a university-scale facility.
The two field-period, QA configuration features a double null-like NRD beyond the aspect ratio 6.6 surface (see \Cref{fig:designA}). Diversion is provided by X-points at the top and bottom of the device (with rotational transform $\iota=0$), which do not form magnetic island chains (referred to as unpaired X-points in~\cite{davies_topology_2025}), topologically similar to the poloidal divertor of tokamaks  \citep{soboleva1997energy, stangeby2000tutorial}. Recently,  QA configurations with poloidal divertors have been recognized in studies of commercial reactors \citep{Thea_2025}, increasing the relevance and potential impact of STAR\_Lite. 

To produce a wide-range of magnetic configurations and satisfy \ref{goal:perturb}, \textbf{design A} is optimized for experimental flexibility. By varying coil currents, a range of QA magnetic configurations can be produced, all of which preserve the NRD. The NRDs not only persist across these configurations, but numerical strike-line simulations confirm that heat exhaust remains resilient to changes in coil currents. In addition, numerical experiments demonstrate that the NRD in \textbf{design A} is robust to coil manufacturing and placement errors, as well as increases in the plasma current.  
To minimize manufacturing complexity and cost without compromising experimental fidelity, \textbf{design A} uses only two distinct coil geometries. The coils will be wound with a ``spine-based'' winding technique (discussed in \Cref{sec:design_criteria}), which reduces the need for high-precision milling and lowers fabrication costs. This approach allows the coils to be wound by students, who will be directly involved in building and operating the device.

The paper is structured as follows.
We motivate the fundamental criteria guiding design decisions in \Cref{sec:design_criteria}. In \Cref{sec:design_process}, we describe the optimization procedure, detailing the evolution from the initial equilibrium to the flexible, simplified coil set of \textbf{design A}. \Cref{sec:performance_analysis} presents a comprehensive analysis of the configuration's physics properties, including magnetic topology, particle confinement, and divertor performance. In \Cref{sec:sensitivity}, we assess the configuration's resilience to manufacturing errors, finite beta, and changes in plasma current. We conclude with a summary and outlook on the future of the STAR\_Lite experiment in \Cref{sec:outlook}.

\section{Key design criteria}
\label{sec:design_criteria}
The design of the first STAR\_Lite coil configuration is driven by three broad requirements: to achieve \ref{goal:build}, the magnetic field must possess an NRD structure; to achieve \ref{goal:measurement}, the design must be practically buildable and operable within the means of the facility; to achieve \ref{goal:perturb}, the design must have sufficient experimental flexibility to produce multiple distinct magnetic geometries. 
This section motivates the specific design criteria used to find a satisfactory stellarator design (summarized in \Cref{tab:design_criteria}). 

\setlength{\tabcolsep}{10pt}
\begin{table}
  \begin{center}
    \def~{\hphantom{0}}
    \begin{tabular}{llll}
      \hline
      \textbf{Parameter} & \textbf{Value} & \textbf{Parameter} & \textbf{Value} \\
      \hline
      Major radius ($R_0$) & 0.5\,\unit{\meter} & Mean field strength on axis ($B_0$) & 0.0875\,\unit{\tesla} \\
      Minor radius ($a$) & 0.075\,\unit{\meter} & Max coil current $(I_{\max})$& 60\,\unit{\kilo\ampere\cdot\mathrm{turns}} \\
      Aspect ratio of opt. surf. & $\sim$ 6.6 & Coil turns $(n_{\text{turns}})$ & 18 \\
      Quasisymmetry & QA & Coil length ($L_{\max}$) & $3$\,\unit{\meter} \\
      Coil-surface dist. ($d_{\text{cs}}$) & 0.06\,\unit{\meter} & Coil-coil dist.  ($d_{\text{cc}}$)  & 0.15\,\unit{\meter} \\
      Max curvature ($\kappa_{\max}$)  & $10.4\,\unit{\meter}^{\sm 1}$ & Mean-squared curvature ($\kappa_{\text{msc}}$)  & $20.0\,\unit{\meter}^{\sm 2}$ \\
      \hline
    \end{tabular}
    \caption{Principal criteria guiding \textbf{design A}, in addition to the existence of stable X-point.}
    \label{tab:design_criteria}
  \end{center}
\end{table}

Building on previous work on QA stellarator design \citep{giuliani_comprehensive_2025} and $\iota=0$ X-point studies \citep{davies_topology_2025}, we restrict the field to be (approximately) QA. 
The remaining criteria on the size, scale, and complexity of the design are determined by practical limitations due to funding, timing, space, available materials, and operational limitations. 

The scale of the device is determined by the space and supply availability at Hampton University. To fit the experiment on an existing $2\,\unit{\meter}\times 3\,\unit{\meter}$ optics table within the HU Fusion Lab, the plasma major radius must be within $R_0=0.5\,\mathrm{m}$. Similarly, the on-axis magnetic field strength is bound by the available heating hardware: two $15\,\unit{\kilo\watt}$ magnetron systems.
A $B_0 = 87.5\,\mathrm{mT}$ magnetic field is required for the magnetron's $2.45\,\unit{\giga\hertz}$ microwaves to resonate with the electron cyclotron frequency. Unlike the major radius, the aspect ratio is not constrained by space requirements. However, once an initial candidate design was found (device \texttt{0104183} from the QUASR database, \href{https://quasr.flatironinstitute.org/model/0104183}{[link]}) the aspect ratio of the optimization surface is kept fixed during the design process to $AR = 6.66$ (discussed more in \Cref{sec:design_process}).

\begin{figure}
    \centering
    \includegraphics[width=0.5\linewidth]{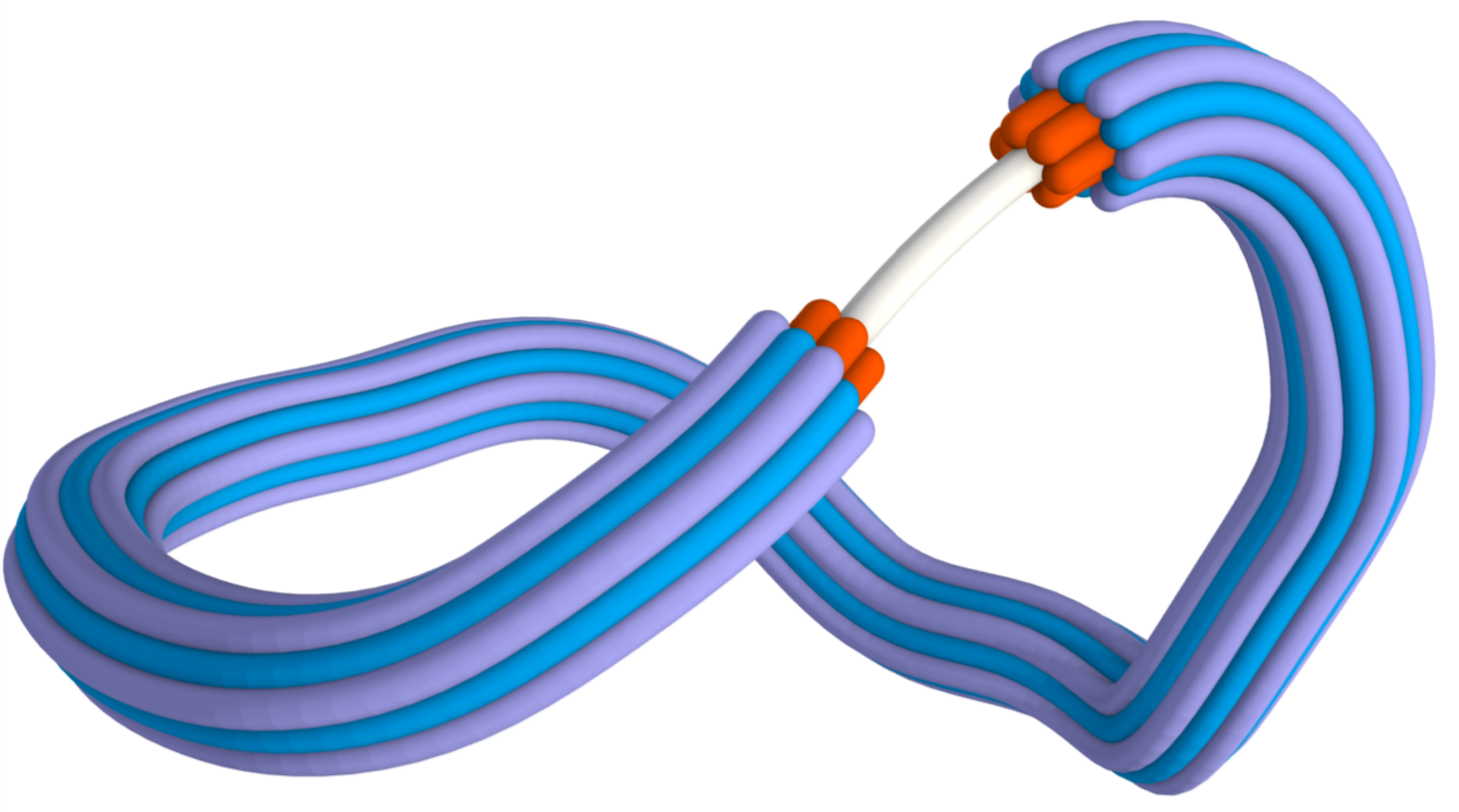}
    \hspace{1cm}
    \includegraphics[width=0.35\linewidth]{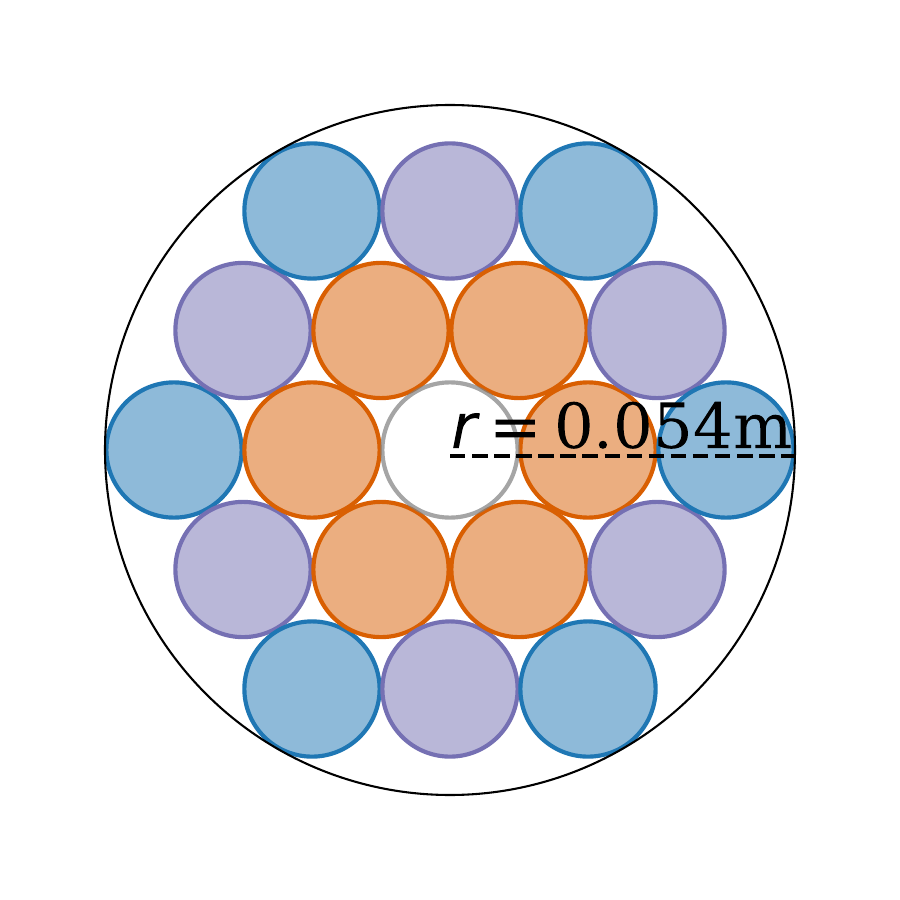}
    \caption{(Left) Visualization of the copper cable (orange, blue, purple) wrapped around the stainless steel central spine (silver), demonstrating ``spine-based'' coil winding. (Right) A cross-section of the winding pack showing three layers of copper cable (orange, blue, purple) forming a hexagonal lattice around the center spine (silver).
    }
    \label{fig:coil_spine_design}
\end{figure}

A central challenge for university-scale stellarators is minimizing manufacturing complexity, manufacturing cost, and operational cost without compromising experimental fidelity. 
To reduce costs and complexity, the coils are designed with a ``spine-based'' winding technique in mind, where current carrying conductors are guided along a stainless steel spine (shown in \Cref{fig:coil_spine_design}). The spine is formed using a free-form 6-axis bending machine and welded to match the precise coil curvature of the design. The limits of the bending machine places a loose constraint on the maximum curvature of the coils, $45.45\,\unit{\meter}^{\sm 1}$; the final coil set had a maximum curvature, $\kappa_{\max} = 10.4\,\unit{\meter}^{\sm 1}$, and mean-squared curvature, $\kappa_{\text{msc}} = 20.0\,\unit{\meter}^{\sm 2}$. The length of each spine is targeted to be $L_{\max} = 3\,\unit{\meter}$.
The low magnetic field strength enables the use of copper cable instead of high temperature superconducting tape, which in turn allows for ambient or water-cooled coils in place of cryogenics, and reduces manufacturing costs incurred during milling \citep{gil_manufacturing_2025} and operational costs. In addition, the coil can be wound by students, who will be building and operating the device.

Electrical and thermal constraints further define the coil cross-section. Preliminary estimates indicate that generating the target on-axis field requires roughly $50\,\unit{\kilo\ampere}\cdot\mathrm{turns}$ of coil current -- within the $I_{\max} = 60\,\unit{\kilo\ampere}\cdot\text{turn}$ limit of available power supplies. 
To balance compactness with cooling, a high-current, low-turn-count design is selected with 18 turns of (4/0) American wire gauge, silicone-insulated, copper welding cable carrying a current of up to $3\,\unit{\kilo\ampere}$. The conductor is wound in a three-layer hexagonal lattice around the spine, forming an $a_{\text{coil}} = 54\,\unit{\milli\meter}$ minor radius winding pack (visualized in \Cref{fig:coil_spine_design} (right)). The diameter of the winding pack, requires the coils spines to be at least $d_{\text{cc}} = 0.15\,\unit{\meter}$ apart (roughly $3\times a_{\text{coil}}$), and the coils and plasma must be at least $d_{\text{cs}} = 0.06\,\unit{\meter}$ apart. 
A simplified adiabatic estimation suggests that the winding pack can sustain a full current pulse for approximately $70\,\unit{\second}$ before experiencing a temperature rise of $100\,\unit{\degreeCelsius}$, keeping the temperature well under the $200\,\unit{\degreeCelsius}$ material tolerance, and leaving open the possibility for steady state operation at reduced current.

The last requirement of the design is the flexibility to experimentally test the resilience of the NRD to magnetic perturbations in a range of magnetic geometries. This is accomplished by designing a coil set with multiple independent current groups; by varying the ratio between the current groups, the magnetic topology can be manipulated. Variable coil currents have been used to control magnetic topology before, for instance in HSX \citep{gerhardt_measurements_2004}. In \Cref{sec:design_process}, the coil set is optimized to produce multiple QA geometries with an X-point, across a wide range of rotational transform profiles, simply by varying the current ratios. Supporting multiple \textit{optimized} magnetic configurations, enables measurement of a range of divertor geometries and the response to changing plasma parameters.

\section{Optimization of \textbf{design A}}
\label{sec:design_process}

The design of STAR\_Lite began with \href{https://quasr.flatironinstitute.org/model/0104183}{QUASR configuration \texttt{0104183}}~\citep{giuliani_comprehensive_2025,giuliani_direct_2024}. The QUASR configuration became a natural candidate after \cite{davies_topology_2025} discovered the unpaired X-point structure beyond the volume of nested magnetic surfaces. In addition to satisfying the divertor criterion, the $\nfp=1$ stellarator symmetric device exhibits excellent quasi-axisymmetry ($0.16\%$ violation) and has only $\ncoils = 6$ modular coils with plenty of room for plasma access. Despite the good qualities of QUASR-\texttt{0104183}, it does not meet all design criteria. The manufacturing complexity of the device is higher than necessary given the three distinct coil geometries, and experimental flexibility is limited since we find that quasisymmetry quality degrades significantly upon changing the coil current ratios (see \Cref{fig:qs}). 
This prompted a redesign of the QUASR-\texttt{0104183} to improve quasisymmetry and experimental flexibility, which we describe in the remainder of this section.

\subsection{Simplifying the initial coil configuration}
To simplify QUASR-\texttt{0104183}, we begin by reducing the number of distinct coil geometries. 
 In addition to exactly satisfying stellarator symmetry, the QUASR-\texttt{0104183} coil set approximately satisfies two-field-period symmetry. This observation allows simplification of the device from three independent coil geometries to two, by imposing two-fold rotational symmetry exactly. QA devices with one-field period symmetry that resemble devices with two-field period symmetry have been observed before by \citep{Landreman_2022}. 

In QUASR-\texttt{0104183}, two of the six original coils are approximately stellarator symmetric, while the remaining four coils are all approximately identical after stellarator and two-fold rotational symmetry applied. 
We call a coil ``stellarator-symmetric'' if its coordinates satisfy $x(t)=x(\sm t), y(t)=\sm y(\sm t)$ and $z(t)=\sm z(\sm t)$, or, equivalently, if the Fourier representation of $x$ comprise only cosine harmonics, while $y, z$ comprise only sine harmonics. The concept of stellarator symmetric coils has also been adopted for the Columbia Stellarator Experiment (CSX) redesign \citep{Baillod_2025}. 
Removing the stellarator-symmetry breaking harmonics from the approximately stellarator-symmetric coil results in the $T$-coil (\Cref{fig:designA}).
The $L$-coils (\Cref{fig:designA}) are obtained by retaining only one of the original coil geometries and applying symmetries to obtain the other three.

The symmetrized coil set, containing only two distinct coils, is polished with a FOCUS-like stage-II optimization to ensure the resulting coil system has nested surfaces and retains the divertor structure \citep{zhu_new_2017}.
The FOCUS-like optimization minimizes the quadratic flux between the coils and the boundary surface of QUASR-\texttt{0104183}, with the rotational and stellarator-symmetry-break harmonics removed.
We observe that the X-point topology is preserved under the coil symmetrization, though the quality of quasisymmetry is degraded by more than an order of magnitude (see \Cref{fig:qs}(a)).

\subsection{Direct coil re-optimization for experimental flexibility} 
\label{sec:direct_optimization}
\ref{goal:perturb} of the STAR\_Lite experiment is to experimentally test the resilience of the NRD structure to magnetic perturbations, in a range of magnetic geometries.
We achieve experimental flexibility by re-optimizing the coil geometries, so distinct QA configurations can be attained by only varying the current ratios between the $T$ and $L$ coils.
We now describe the partial differential equation (PDE)-constrained minimization problem solved to find the first STAR\_Lite coil configuration. 

The optimization problem maintains degrees of freedom, $\cb = (\cb_1, \cb_2)$, for the two distinct coil geometries ($L$ and $T$ coils), as well as degrees of freedom for three distinct groups of coil currents, $\Ib = (\Ib_1, \Ib_2, \Ib_3)$. The optimization seeks coil geometries such that for each current group, the device is QA on a surface with aspect ratio approximately 6.6 and that surface, called optimization surface, is constrained to have a specific rotational transform $\iota_1  = 0.13, ~ \iota_2  = 0.18, ~\iota_3  = 0.26$. We solve the following problem:
\begin{subequations}
\begin{align}
    \min_{ \mathbf c, \mathbf{I}, \mathbf s} ~& f(\mathbf c, \mathbf{I}, \mathbf s) \notag
    \\
    \text{s.t. } \mathbf r( \mathbf c, \mathbf{I}_k, \mathbf s_k) &= 0 
    \label{eq:bsurf_constraint} 
    \\
    \text{Vol}(\mathbf{s}_k) &= V, 
    \label{eq:volume_constraint}
    \\
    \iota(\mathbf{s}_k) &= \iota_k, 
    \label{eq:iota_constraint}
    \\
    R(\mathbf s_2) &= R_0, \label{eq:major_radius_constraint}
    \\
    \|\bm \Gamma_{\text{coil},i}(\cb) - \bm \Gamma_{\text{coil},j}(\cb)  \|_2 &\geq d_{\text{cc}}, \label{eq:cc_dist_constraint}
    \\
    \|\bm \Gamma_{\text{surface}}(\mathbf s_k) - \bm \Gamma_{\text{coil},i}(\cb) \|_2 &\geq d_{cs} , \label{eq:cs_dist_constraint}
    \\
    \text{Length}_i(\cb) &\leq L_{\max} ,
    \label{eq:length_constraint}
    \\
    \text{Curvature}_i(\cb) &\leq \kappa_{\max} ,
    \label{eq:max_curvature_constraint}
    \\
    \text{Mean-Squared Curvature}_i(\cb) &\leq \kappa_{\text{msc}},
    \label{eq:msc_constraint}
    \\
    \| \Ib_k \|_\infty &\leq I_{\max} , \label{eq:current_constraint}
\end{align} \label{eq:optimization_problem}
\end{subequations}
for $k=1,2,3$ and $i,j=1,\ldots, \ncoils$.
The vectors $\mathbf{s}_1,\mathbf{s}_2,\mathbf{s}_3$ denote the degrees of freedom of the optimization surfaces on which quasisymmetry is targeted in the three different configurations, respectively. 
Surface, $\mathbf{s}_k$, is computed from the coil geometries and currents $\Ib_k$ using the Boozer surface technique from \citep{giuliani_direct_2022}, in order to enforce the PDE constraint exactly, \cref{eq:bsurf_constraint}. 
The three surfaces are all constrained to have the same volume as QUASR-\texttt{0104183} through \cref{eq:volume_constraint}.
As shown in \citep{giuliani_direct_2022}, computing the magnetic surfaces facilitates computation of the quasisymmetry metrics and the rotational transform $\iota(\mathbf{s}_k)$.

The objective  
\begin{equation}
   f(\mathbf c, \mathbf{I},\mathbf s) = \frac{1}{3}\sum_{k=1}^3 f_{QA}(\mathbf c, \mathbf I_k, \mathbf s_k)
\end{equation}
measures the average violation from quasisymmetry of the three configurations, where the individual quasisymmetry violation is
\begin{equation}\label{eq:qa1}
f_{QA}(\mathbf c, \mathbf I_k, \mathbf s_k) = \frac{\iint_S(B(\varphi, \theta;\mathbf c, \mathbf I_k, \mathbf s_k) - B_{\text{QA}}(\theta;\mathbf c, \mathbf I_k, \mathbf s_k) )^2  ~dS}{\iint_S B_{\text{QA}}(\varphi, \theta;\mathbf c, \mathbf I_k, \mathbf s_k)^2 ~dS},
\end{equation}
and
\begin{equation}\label{eq:qa2}
B_{\text{QA}}(\theta;\mathbf c, \mathbf I_k, \mathbf s_k) = \frac{\iint_S B(\varphi,\theta;\mathbf c, \mathbf I_k, \mathbf s_k)~dS}{\iint_S ~dS}
\end{equation}
is the closest quasisymmetric field strength, in a least squares sense, to the true one generated by the coils.

\Cref{eq:iota_constraint} requires that the on-surface rotational transform in the three configurations is equal to the target values $\iota_1, \iota_2, \iota_3$.
\Cref{eq:major_radius_constraint} reflects the design criteria that the major radius of the device is $R_0$ when current $\Ib_2$ is applied. The major radius is not constrained in the other two configurations since it does not vary substantially, given the volume constraints.
\Cref{eq:cc_dist_constraint} requires that the coil-to-coil distance is greater than $d_{\text{cc}}$. The notation $\Gammab_{\text{coil},i}(\cb), \Gammab_{\text{surface}}(\mathbf{s}_k)$ denotes the position vector of a point on coil $i$ and surface $k$; the dependence on an arc length parameter which exactly define the position is neglected for brevity.
\Cref{eq:cs_dist_constraint} requires that the coil-to-surface distance be at least $d_{cs}$.
Constraint \cref{eq:length_constraint} requires the length of the coils be at most $L_{\max}$; this is the approximate length of the coils in QUASR-\texttt{0104183}. Constraints \cref{eq:max_curvature_constraint,eq:msc_constraint} requires the pointwise curvature and mean-squared curvature of the coils be at most $\kappa_{\max},\kappa_{\text{msc}}$; these are the approximate curvature bounds on the coils in QUASR-\texttt{0104183}. 
Constraint \cref{eq:current_constraint} requires that that coil currents not exceed $I_{\max}$, reflecting manufacturing constraints discussed in \Cref{sec:design_criteria}.

The optimization problem, \cref{eq:optimization_problem}, is solved by eliminating the degrees of freedom associated with the surfaces $\mathbf s_k$ by enforcing a PDE constraint in \cref{eq:bsurf_constraint}, and volume constraints, \cref{eq:volume_constraint}, exactly. 
The eliminated surface degrees of freedom correspond to the Fourier harmonics of the magnetic surface, parametrized in Boozer coordinates.
The reduced objective $\hat f(\mathbf c, \Ib) = f(\mathbf c, \Ib, \mathbf s(\mathbf c, \Ib))$ is minimized using the L-BFGS-B minimization algorithms. Discretely exact derivative information is obtained using an adjoint approach discussed in \citep{giuliani_direct_2022}.

The (in)equality constraints \cref{eq:iota_constraint,eq:major_radius_constraint,eq:cc_dist_constraint,eq:cs_dist_constraint,eq:length_constraint} are imposed inexactly using a quadratic penalty method: weighted quadratic penalties are added to the objective, and the corresponding weights are increased until the constraint violation is acceptable \citep{nocedal_numerical_2006}.
The current constraint \eqref{eq:current_constraint} is a bound constraint, which can be enforced exactly using, e.g., the L-BFGS-B minimization algorithm. One current per configuration is held fixed during the optimization to prevent the field strength from going to zero; after the optimization, the currents are rescaled such that the on-axis field strength is $B_0$.
The final STAR\_Lite device, which we refer to as \textbf{design A}, shown in \Cref{fig:designA}, is visually similar to the starting configuration QUASR-\texttt{0104183} but meets the design criteria in \Cref{sec:design_criteria}.

\section{Design A performance, magnetic topology, and divertor analysis}
\label{sec:performance_analysis}

We now investigate the physics properties of the final design produced by the direct re-optimization. The primary goals of this analysis are twofold: demonstrate that the configuration meets the design criteria outlined in \Cref{sec:design_criteria}, and ensure that the divertor structure is present and will meet the experimental goals. Some basic geometric properties of the three configurations for which the coilset is optimized (low, medium, and high $\iota$) are shown in \Cref{tab:geometric_properties}.
\begin{table}
    \centering
    \caption{Geometric properties of the three primary \textbf{design A} configurations. The effective minor radius, plasma volume, surface area, and aspect ratio, are reported on the last closed flux surface (not to be confused with the surface targeted in optimization).}
    \label{tab:geometric_properties}
    \begin{tabular}{lcccccc}
        \hline
        \textbf{Config.} & \begin{tabular}{@{}c@{}}Effective \\ minor radius [m]\end{tabular} & \begin{tabular}{@{}c@{}}Plasma \\ Volume [liters]\end{tabular} & \begin{tabular}{@{}c@{}}Surface \\ Area [$\mathrm{m}^2$]\end{tabular} & Aspect Ratio \\
        \hline
        \colorrect{iotaLow} Low-$\iota$   & 0.08 & 67  & 2.18 & 6.2 \\
        \colorrect{iotaMedium} Medium-$\iota$ & 0.10 & 96  & 2.55 & 5.1 \\
        \colorrect{iotaHigh} High-$\iota$    & 0.13 & 166 & 3.32 & 3.8 \\
        \hline
    \end{tabular}
\end{table}
These analyses are performed using a standard filamentary coil approximation. To verify the validity of this assumption, we compare with predictions using a finite-build coil model which consists of 18 filamentary winds in a hexagonal winding structure shown in \Cref{fig:coil_spine_design}. The difference between the model predictions is surprisingly small: the finite-build structure changes the quasisymmetry violation by roughly $0.2\%$, the mean field strength on axis by less than 1\,$\mathrm{\mu T}$, and moves the X-point by at most $1.1\,\unit{\milli\meter}$.

\subsection{Quasisymmetry, configuration flexibility, and particle loss estimates}
\begin{figure}
    \centering
    \includegraphics[width=\linewidth]{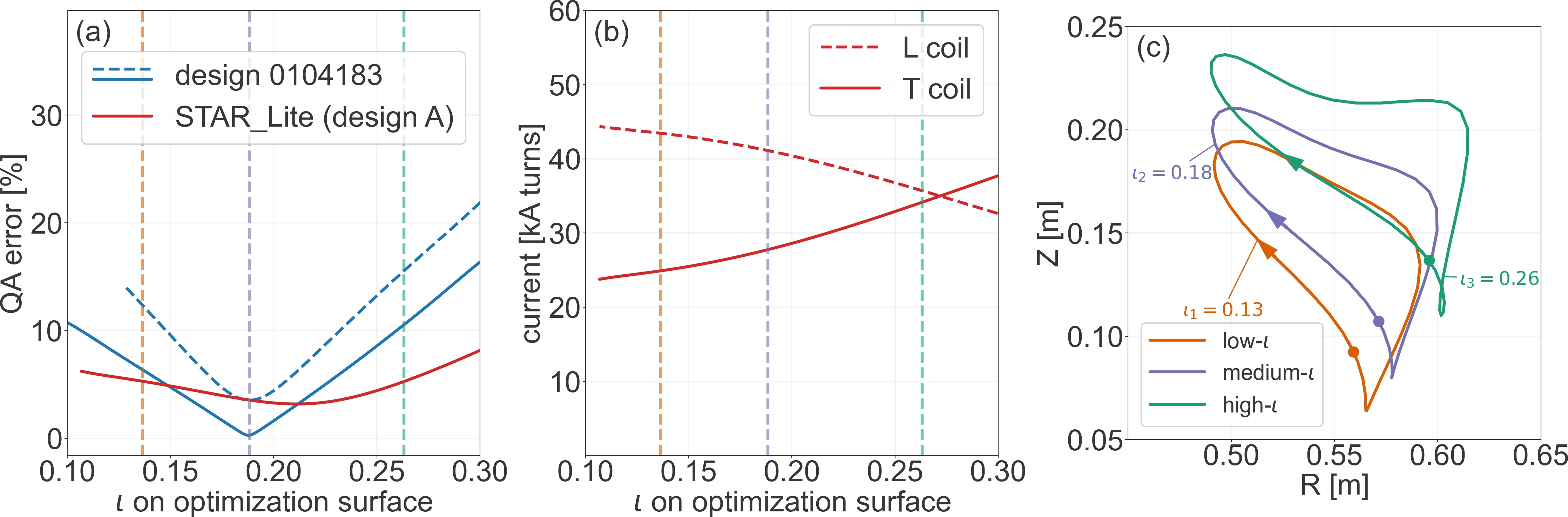}
    \caption{a) The quasisymmetry error as a function of the rotational transform on the optimized surface, obtained by varying the current ratio in $L$ and $T$-coils. 
    The solid and dashed blue lines correspond to QUASR-0104183 before and after the symmetrization procedure, respectively.
    b) The currents in the $L$ and $T$-coils as a function of the rotational transform on the optimized surface. For each device, the field strength on the magnetic axis is $B_0$.
    In panels a) and b), the vertical dashed lines correspond to the rotational transforms targeted on the optimization surface.
    c) The $(R, Z)$ positions of the upper X-points for the three $\iota_k$-configurations as the cylindrical angle $\phi$ is varied from $0$ to $\pi$.  The dot indicates cylindrical $\phi=0$, and $\phi$ increases in the clockwise direction, indicated by the arrow.
    }
    \label{fig:qs}
\end{figure}%
The performance of the design produced by the direct re-optimization is demonstrated in \Cref{fig:qs}.
\Cref{fig:qs}(a) shows the quasisymmetry error as a function of the rotational transform on the optimization surface, obtained by varying the current ratio in the $L$ and $T$-coils. The three curves represent the original QUASR-\texttt{0104183} device (solid, blue line), its symmetrized counterpart (dashed, blue line), and \textbf{design A} (solid, red line). 
The vertical dashed lines correspond to the values of rotational transform targeted in the optimization problem, given by \cref{eq:optimization_problem}. \Cref{fig:qs}(a) shows that the direct optimization reduced the quasisymmetry error by over 50\% for the low and high-iota configurations, without any increase in quasisymmetry error at the middle configuration ($\iota_2$), relative to the symmetrized device.  
To generate the quasisymmetry error curve in \Cref{fig:qs}(a) for the QUASR-\texttt{0104183} device, we uniformly scale the original \textit T-coil and \textit L-coil currents as a group.

Compared to the QUASR-\texttt{0104183} device, \textbf{design A} exchanges QS quality at the middle configuration ($\iota_2$), for QS quality at the low and high rotational transform configurations.
\textbf{Design A} has quasisymmetry error that varies much less with respect to the rotational transform: approximately $3.5\%$ QA error for the middle configuration, and approximately $5\%$ error for both the low and high-iota configurations, respectively. 
Achieving sub 5\% QS-error across a range of rotational transform variables, \textbf{design A} demonstrates experimental flexibility \ref{goal:perturb}.
\Cref{fig:qs}(b) shows the currents in the $L$ and $T$ coils in \textbf{design A} used to generate the corresponding rotational transform. %
\Cref{fig:qs}(c) shows the cylindrical $(R, Z)$ positions of the upper X-points for the three $\iota_k$-configurations when $\phi \in [0, \pi/2]$. The cylindrical angle, $\phi$, increases in the clockwise direction from the dot, where $\phi=0$.

We now evaluate the confinement capability of the  configurations.
To quantify particle confinement in \textbf{design A}, we trace collisionless guiding center trajectories of 1,000 electrons spawned on the magnetic axis until a final time of $0.02\,\unit{\second}$ with two energies: a standard $20\,\unit{\electronvolt}$ thermal electron; a $2.86\,\unit{\kilo\electronvolt}$ \say{fast-electron}. 
A particle is considered lost if it leaves the $AR=6.6$ optimization surface on which quasisymmetry is targeted.
Tracing standard electrons brings insight into the confinement time expected for electrons in experiments, whereas tracing fast-electrons emulates fast-ions in a reactor-scale design. The energy of the fast-electron is selected so that the ratio of the Larmor radius to minor radius matches that of a fast-ion in a reactor-scale design using
 \begin{equation}
    E_e = E_\alpha\frac{ m_\alpha Z_e^2 B_0^2 a^2}{m_e Z_\alpha^2B_\alpha^2 a_\alpha^2} = 2.86 \,\unit{\kilo\electronvolt},
\end{equation}
with electrons ($Z_e =\sm 1, m_e = 9.1\times 10^{\sm 31}\,\unit{\kilo\gram}$) in a device with a mean field strength on axis of $B_0 = 100$\,mT and a minor radius, $a = 0.075$\,m, at birth emulate $E_\alpha = 3.52\,\unit{\mega\electronvolt}$ $\alpha$-particles ($Z_\alpha = 2, m_\alpha= 6.7\times 10^{\sm 27}\,\unit{\kilo\gram}$) moving through an Infinity-Two \citep{hegna_infinity_2025} scale device ($B_\alpha=9\,\unit{\tesla}$ average field strength, $a_\alpha = 1.25\,\unit{\meter}$ minor radius). 
Electrons can be launched at this energy with an electron gun.

\begin{figure}
    \centering
    \includegraphics[width=\linewidth]{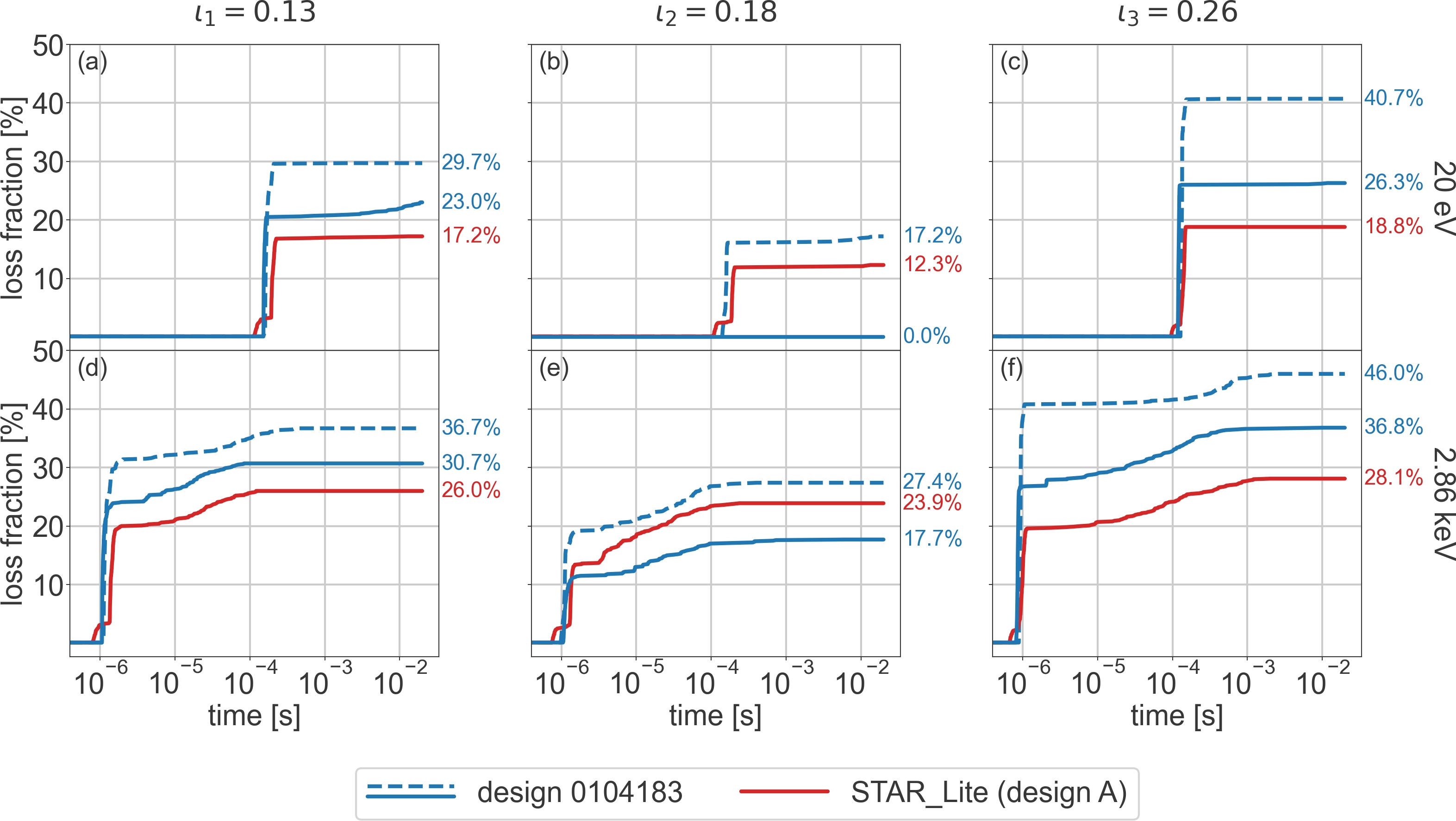}
    \caption{Loss fraction as a function of time for two electron energies ($20\,\mathrm{eV}$ and $2.86\,\mathrm{keV}$) for the three iota configurations. 
    The solid and dashed blue lines correspond respectively to the original QUASR-\texttt{0104183} design and its symmetrized counterpart, while the solid red line corresponds to \textbf{design A}.}
    \label{fig:losses}
\end{figure}

\Cref{fig:losses} shows the loss fractions with respect to time, for each of the three rotational transform configurations (columns) and two electron energies (rows).
The proportion of particles lost at the final time behaves similarly to the quasisymmetry error in \Cref{fig:qs}(a).
Once again, \textbf{design A} (solid red line) improves particle confinement at all $\iota$ values compared to the symmetrized QUASR device (dashed blue line), and improves particle confinement at the high and low $\iota$ configurations relative to the original QUASR device (solid blue line).
The electron confinement time and loss fraction is acceptable to map the magnetic field, addressing \ref{goal:measurement}.

\subsection{Magnetic topology}
We analyze topological aspects of the three targeted configurations: the rotational transform ($\iota$) profile, the size/shape of magnetic flux surfaces, hyperbolic fixed points (referred to as the X-points),  Greene's residue of X-points, and the invariant manifolds associated with the X-points that form the separatrix and divertor legs. The X-point and the associated invariant manifolds are calculated using a field line map analysis method implemented in \texttt{pyoculus} (\url{https://github.com/akitzu/pyoculus}). 
\begin{figure}
    \centering
    \includegraphics[width=0.9\linewidth]{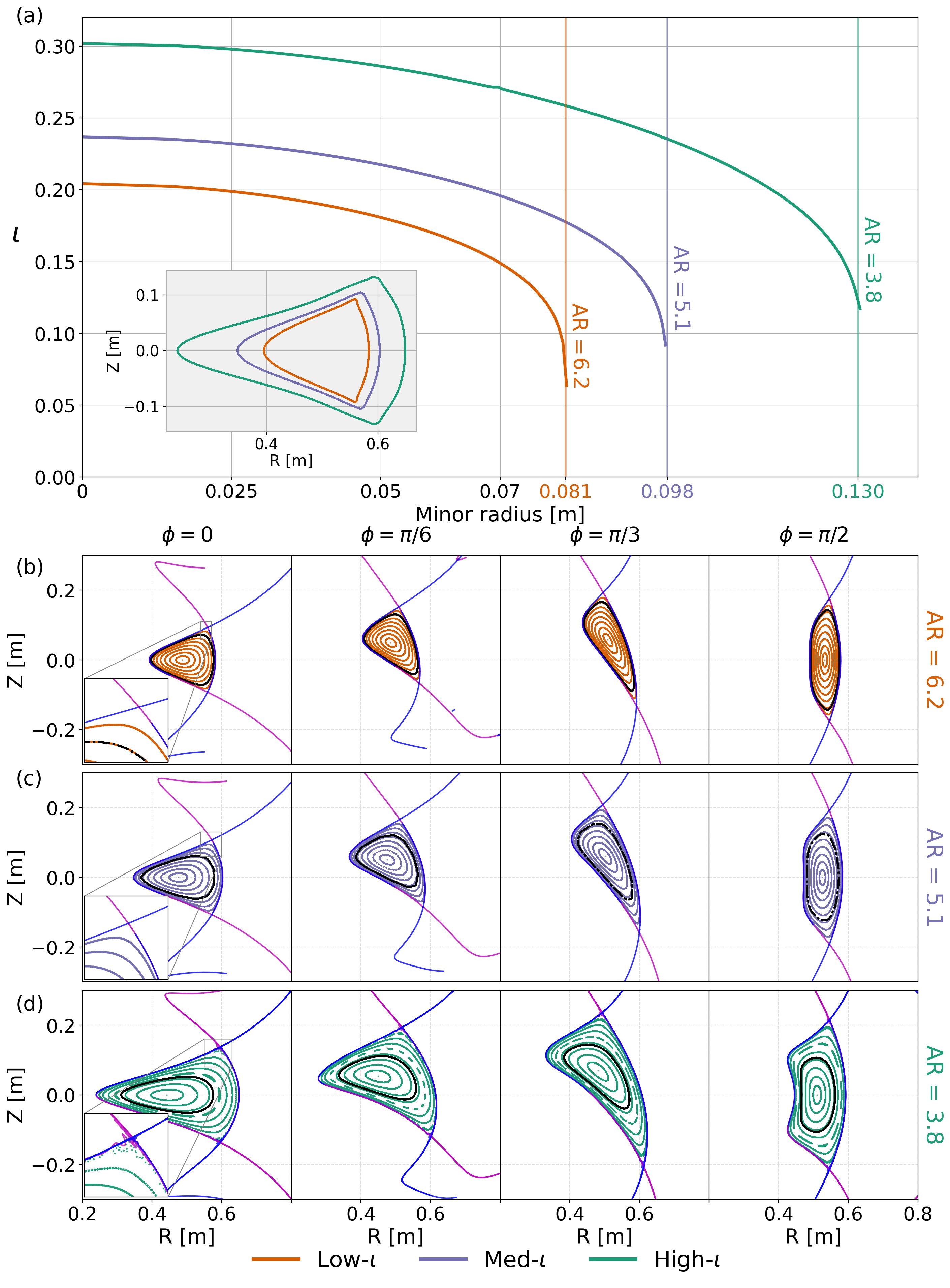}
    \caption{ (a) Rotational transform profiles for the three \textbf{design A} configurations all the way out to the last closed flux surface, indicated by vertical lines. The inset shows the shape of the last closed flux surface.
    (b) - (d) Poincar{\'e} sections and invariant manifolds associated with the X-points, for the three \textbf{design A} configurations (rows), at four values of $\phi$ (columns). The black curve on the Poincaré sections denotes the optimization surface on which the rotational transform is targeted.
    }
    \label{fig:manifolds_v2}
\end{figure}
\Cref{fig:manifolds_v2}(a) presents the rotational transform profiles for the low (\protect\colorrect{iotaLow}), medium (\protect\colorrect{iotaMedium}), and high-$\iota$ (\protect\colorrect{iotaHigh}) configurations, with vertical lines indicating the minor radius of the respective last closed flux surface (shown in the inset) used to calculate the aspect ratio. This demonstrates the different aspect ratios of the magnetic field configurations that can be achieved in \textbf{design A}. 
\Cref{fig:manifolds_v2}(b-d) shows Poincar\'e sections of the three configurations at four cylindrical angles, and the target optimization surface. The Poincar\'e sections reveal that varying the coil currents induces significant changes in aspect ratio, also shown in \Cref{tab:geometric_properties}, and demonstrates that STAR\_Lite can generate 3 distinct stellarator equilibria within a single device. 
Notably, a large volume of \textit{nested} flux surfaces is preserved under the change in coil currents. 
Additionally (not shown here), we have found that coil currents can be varied to produce another configuration with nested flux surfaces, X-points, and an $\iota$-on-optimization-surface as large as 0.4.

\Cref{fig:manifolds_v2}(b-d) also shows the X-points -- non-axisymmetric, closed field lines with zero rotational transform -- as well as the unstable and stable invariant manifolds associated with the X-point on the same poloidal plane. The invariant manifolds form the set of all magnetic field lines that asymptotically move away from the X-point along the stable or unstable manifold when followed in either the negative or positive toroidal direction, respectively.
Due to the invariance and orientation-preserving nature of the manifolds, they provide topological separation between fieldlines inside and outside the LCFS. In fact, the manifolds as barriers need an in-depth study of lobes and their iterates \citep{meiss2015thirty, smiet_turnstiles_2025} which is beyond the scope of this article.
From each X-point emerge four curves, 2 branches for each of the invariant manifolds. Two of these curves -- the so-called divertor legs -- are formed by the unstable and stable manifold and carry field lines away from the X-point towards the wall when followed in the positive and negative toroidal direction, respectively. The remaining two curves bound the confined plasma and form the last closed flux surface, in the absence of chaos.
\Cref{fig:manifolds_v2}(b,c) shows that in the low and medium-$\iota$ cases, the manifold branches enclosing the plasma overlap completely, forming a separatrix and implying the non-existence of chaotic trajectories in these configurations. 
However, in the high-$\iota$ case, these manifold branches do not completely overlap, and generate lobes as shown in the inset of \Cref{fig:manifolds_v2}(d) -- which proves the existence of chaos due to Smale-Birkhoff homoclinic theorem. A detailed analysis of the lobe geometry and associated heteroclinic tangle dynamics is needed to better understand fieldline transport in the high-$\iota$ configuration; this is left for future study. This analysis verifies the existence of the X-points and mapping the divertor legs shows that \textbf{design A} satisfies the desired NRD structure \ref{goal:build}.

\begin{figure}
    \centering
    \includegraphics[width=\linewidth]{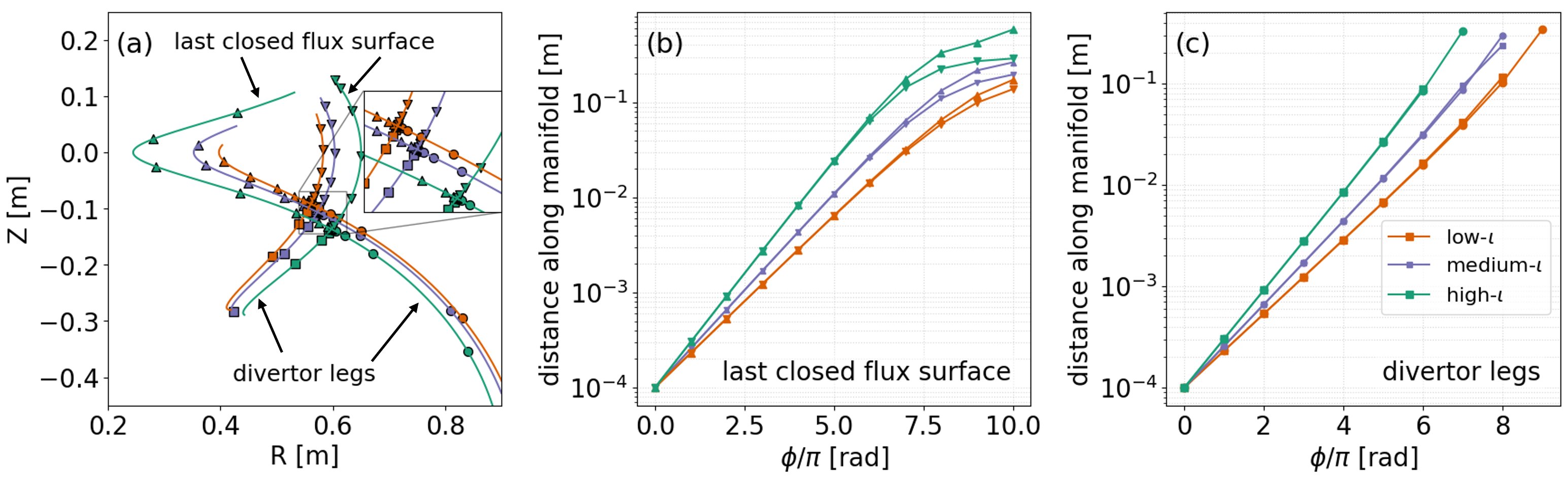}
    \caption{ Illustration of trajectories of field lines initialized $0.1$mm along the bottom X-point manifolds. (a) Colored curves showing the manifolds of the lower X-point at $\phi=0$ for low-$\iota$, medium-$\iota$ and high-$\iota$ configurations, (shown in orange, purple and green respectively). Markers show an individual field line on a given manifold branch every time it intercepts this $\phi=0$ or $\phi=180^\circ$. (b) Showing, for the field lines traveling along the last closed flux surface, the distance traveled along the curve every time $\phi=0$ or $\phi=180^\circ$ is intersected. The x-axis records the toroidal distance traveled by the field line and the distance is measured every $180^\circ$. The X-point located at distance along the curve of $0$. (c) Distance traveled by the selected field lines along the divertor legs each time $\phi=0$ or $\phi=180^\circ$ is intersected. (b) and (c) show that these field lines move along the manifolds faster in the high-$\iota$ configuration which, near to the X-point, is predicted by Green's residue.
    }
    \label{fig:greenes_residue_illustration}
\end{figure}

Finally, we examine the rate at which magnetic field lines travel along the manifolds. The motion of field lines along manifolds determines the plasma transport to leading order, and therefore differences between magnetic configurations are likely to lead to differences in divertor performance. We observe that field lines travel faster along the manifolds in the high $\iota$ case rather than the low $\iota$ case. This is shown explicitly by \Cref{fig:greenes_residue_illustration}, in which we sample a field line on the manifold, $0.1$mm from the X-point for each configuration, and record the distance traveled along the manifold each time the field intersects $\phi=0$ or $\phi=180^\circ$. The curve color indicate which configuration the field line is in, low (\protect\colorrect{iotaLow}), medium (\protect\colorrect{iotaMedium}), and high-$\iota$ (\protect\colorrect{iotaHigh}) configurations. \Cref{fig:greenes_residue_illustration} (a) shows the trajectory of these field lines and \Cref{fig:greenes_residue_illustration} (b) and (c) show the distance traveled along the manifold for selected field lines on the LCFS and divertor legs respectively. 
The field lines traverse the manifolds fastest for the high-$\iota$ configuration. This is described by Greene's residue, a quantity which describes the rate at which field lines traverse the manifolds in the immediate neighborhood of the X-point. The Greene's residue spans the range $\sm0.19$  to $\sm 0.35$, from the low-$\iota$  to high-$\iota$ configuration, with the more negative value indicating faster motion of field lines away from the X-point. Configuration flexibility allows these topological differences to be experimentally compared, which will be a valuable study.

\subsection{Divertor analysis}
\label{sec:physics_analysis}
As described in \Cref{sec:introduction}, we require the edge magnetic field to deposit the exhausted plasma on plasma-facing components in a small number of resilient (to changes in equilibrium), distinct, narrow bands. We examine this by simulating the plasma heat loads on two vessel designs, described in \Cref{sec:vessel_wall}. The heat load analysis suggests placement of divertor plates, although divertor design is left as future work. Furthermore, the studies presented here may provide geometric insights applicable to a scaled-up, reactor-size version of the STAR\_Lite configuration. 

We employ EMC3-Lite \citep{feng2022review}, which uses an anisotropic constant coefficient heat diffusion model,
\begin{equation} 
\frac{\kappa_e}{n_e} \nabla_\parallel^2 T + \chi \nabla_\perp^2 T  = 0. \label{eq:anisotropic_heat}
\end{equation}
The plasma temperature, $T$, is assumed to be the same for ions and electrons. The heat equation is closed by assuming a given net heat flux $P_\text{SOL}$ landing on the PFCs and a Bohm sheath boundary condition \citep{bohm1949characteristics, riemann1991bohm} :
\begin{equation}
    Q_{\parallel, \textrm{PFC}} = -\kappa_e \nabla_\parallel T_\textrm{PFC} = n c_s \gamma T_\textrm{PFC}.\label{eq:bohm_boundary}
\end{equation}
In this model, $\kappa_e = \kappa_{e0}T_\text{ref}^{5/2}$ is the Spitzer heat conductivity \citep{cohen_1950, spitzer_1953} (where $\kappa_{e0}= 1.3\times10^{22}\,\text{eV}^{-5/2}\text{m}^{\sm 1}\text{s}^{\sm 1}$ is a constant), $T_\text{ref}$ is a ``typical'' edge temperature, $n_e$ is the electron density, $\chi$ is the perpendicular thermal diffusivity, $Q_{\parallel, \textrm{PFC}}$ is the parallel heat flux (where $Q$ represents a power per unit area) on the vessel wall, $T_\textrm{PFC}$ is the plasma temperature at the wall, $c_s=\sqrt{2 k_B T_\text{ref}/m_i}$ is the ion sound speed (where $k_B$ is Boltzmann`s constant and $m_i$ is the ion mass) and $\gamma$ is the sheath transmission factor. $\kappa_e$, $n_e$ and $\chi$ (and thus the parallel-to-perpendicular diffusivity strength, $\kappa_{e0}T_\text{ref}^{5/2}/(n_e \chi)$) are taken to be spatially uniform%
. 

EMC3-Lite has been benchmarked against W7-X experimental results in attached conditions \citep{gao_2023}, although it should be emphasized that several physics effects (such as spatially varying transport coefficients, cross-field drifts and neutral and radiation dynamics) are not included. Due to the compact physical size, low absolute density, and low heating power of STAR\_Lite, edge plasma transport will likely be heavily governed by interactions with neutral particles. Higher-fidelity simulations incorporating these interactions are beyond the scope of this initial design paper, and planned for future work.

Here we present results using a parallel-to-perpendicular weight of $\kappa_{e0}T_\text{ref}^{5/2}/(n_e \chi) = 4\times10^{7}$. This is consistent with the parameters ($T_\text{ref}=100$\,eV, $n_e=10^{19}\,\text{m}^{-3}$, $\chi=3\,\text{m}^2\text{s}^{\sm 1}$) used in both W7-X \citep{gao_2023} and as the lower realistic bound of two candidates for stellarator reactors \citep{goodman_squid_2025, davies_squid_2025, lion_stellaris_2025}. It is also consistent with $T_\text{ref}=10$\,eV, $n_e=2\times10^{17}\,\text{m}^{-3}$, $\chi=0.5\,\text{m}^2\text{s}^{\sm 1}$ which for STAR\_Lite probably represents the upper realistic bound of $\kappa_{e0}T^{5/2}/(n_e \chi)$, i.e., the case where the parallel transport is strongest and, correspondingly, the area wetted by the plasma is smallest and the power per unit area greatest. In these simulations, we set the total non-radiated exhausted power ($P_\text{SOL}$) to match the maximum available heating power, $30\,\mathrm{kW}$. %

\subsubsection{Vacuum vessel design}\label{sec:vessel_wall}
The vacuum vessel design is primarily constrained by the necessity of fitting the vessel between the complex non-planar coil geometry and the magnetic field cage. The two candidate vessels examined (\nicefrac{1}{4} segment) are shown on the left side of \Cref{fig:heatflux}.

The first, referred to as the \textbf{extended vessel}, is numerically generated by scaling a surface near the Last Closed Flux Surface (LCFS) of the Medium-$\iota$ configuration by a factor of 3. This design maintains a conformal fit within the coils and avoids intersection with the invariant manifolds for the low and medium-$\iota$ configurations, although it intersects the LCFS of the high-$\iota$ configuration over a small region. Manufacturing such a shape would likely require complex techniques such as casting, hydroforming, additive manufacturing or advanced CNC machining. 

The second, the \textbf{origami vessel}, is designed for simplicity, utilizing only straight faces to ease the integration of ports and internal components. %
This vessel design avoids intersection with the invariant manifolds for all three $\iota$ configurations.  While the faces could be cut and bent from single metal sheets, the high number of required welds presents significant challenges regarding vacuum tightness and potential weld magnetization.

\subsubsection{Heat flux footprints and connection length}
\begin{figure}
    \centering
    \includegraphics[width=\linewidth]{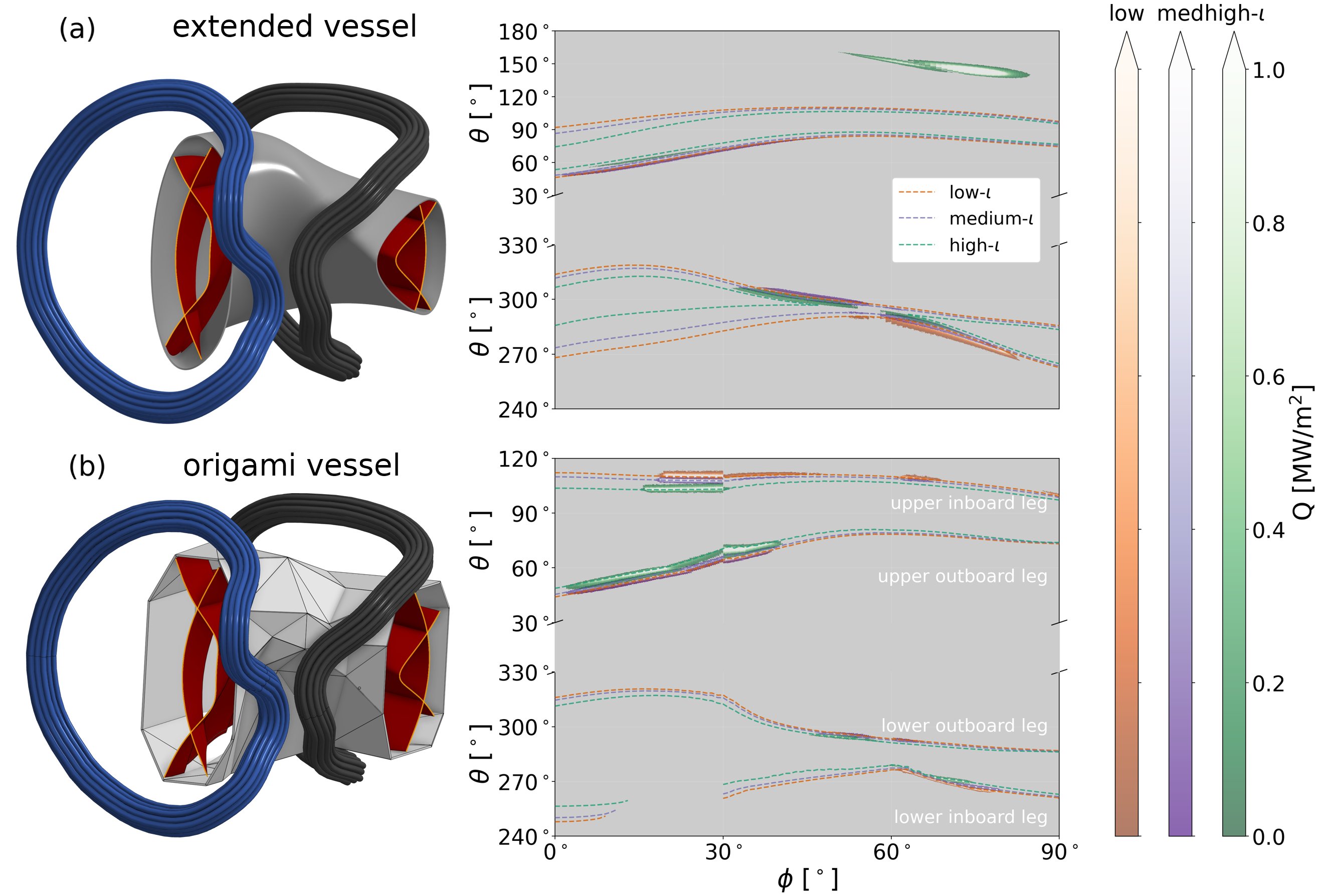}
    \caption{Distribution of simulated wall heat flux for the two vacuum vessel candidates. Left: Three-dimensional views of the magnetic X-point manifolds (red) relative to the L-coil (black), T-coil (blue), and the preliminary vacuum vessel designs (grey). The top row displays the ``extended'' vessel design, while the bottom row shows the ``origami'' vessel design. Right: Heat flux footprints ($Q$ in MW/m²) on the surface of the Extended Vessel and the Origami Vessel, respectively. The dashed lines represent the manifold intersections with the vessel wall for all three iota configurations represented by different colors. The heat flux is plotted as a function of the toroidal angle $\phi$ and poloidal angle $\theta$ for a $90^\circ$ section of the machine. The color scales highlights the strike point patterns, showing a more continuous wetted area in outboard upper and lower divertor of the extended design compared to the more localized strike zones in the origami configuration. 
    }
     \label{fig:heatflux}
\end{figure}

The plasma strike patterns on the vessel walls for the low, medium and high $\iota$ configurations are shown in \Cref{fig:heatflux} as a function of toroidal angle $\phi$ and poloidal angle $\theta$ over a half field period. We use a geometric poloidal angle defined with respect to the magnetic axis of the medium $\iota$ configuration, i.e. $\tan(\theta) = (Z-Z_\text{MA})/(R-R_\text{MA})$, where $(R_\text{MA}(\phi), Z_\text{MA}(\phi))$ is the location of the magnetic axis. 
To guide the eye we also plot the points where the divertor legs intersect the vessel with dashed lines for all $\iota$ configurations. 

For both vessels, the plasma strike pattern falls on discontinuous stripes which wrap around the vessel toroidally%
, coinciding with the intersection of the divertor legs with the vessel. %
Small variations in the divertor leg locations can cause strike location changes of several centimeters, but the heat deposition pattern is relatively stable. This appears to confirm the resilient behavior required of NRDs. It should be noted that, the high-$\iota$ configuration creates an additional hotspot on the extended vessel around $\theta=180^\circ$, the result of the LCFS intersecting the wall over a small toroidal range and creating limiter-like conditions. %
We expect that this issue can be avoided by a more careful vessel design.

It might be surprising that, even though the divertor legs intersect the vessels at all toroidal locations, the heat pattern on the wall is toroidally discontinuous. The reason for this is an effect called target shadowing \citep{feng2022review, kharwandikar2025power}. At certain locations, the 3D geometry of the field and vessel dictates that magnetic field lines on divertor legs are intercepted by the vessel wall before they can reach the X-point. As a result, these field lines cannot be loaded with plasma by parallel transport and result in low heat loads in the so-calledshadowed locations.

\begin{figure}
    \centering
    \includegraphics[width=\linewidth]{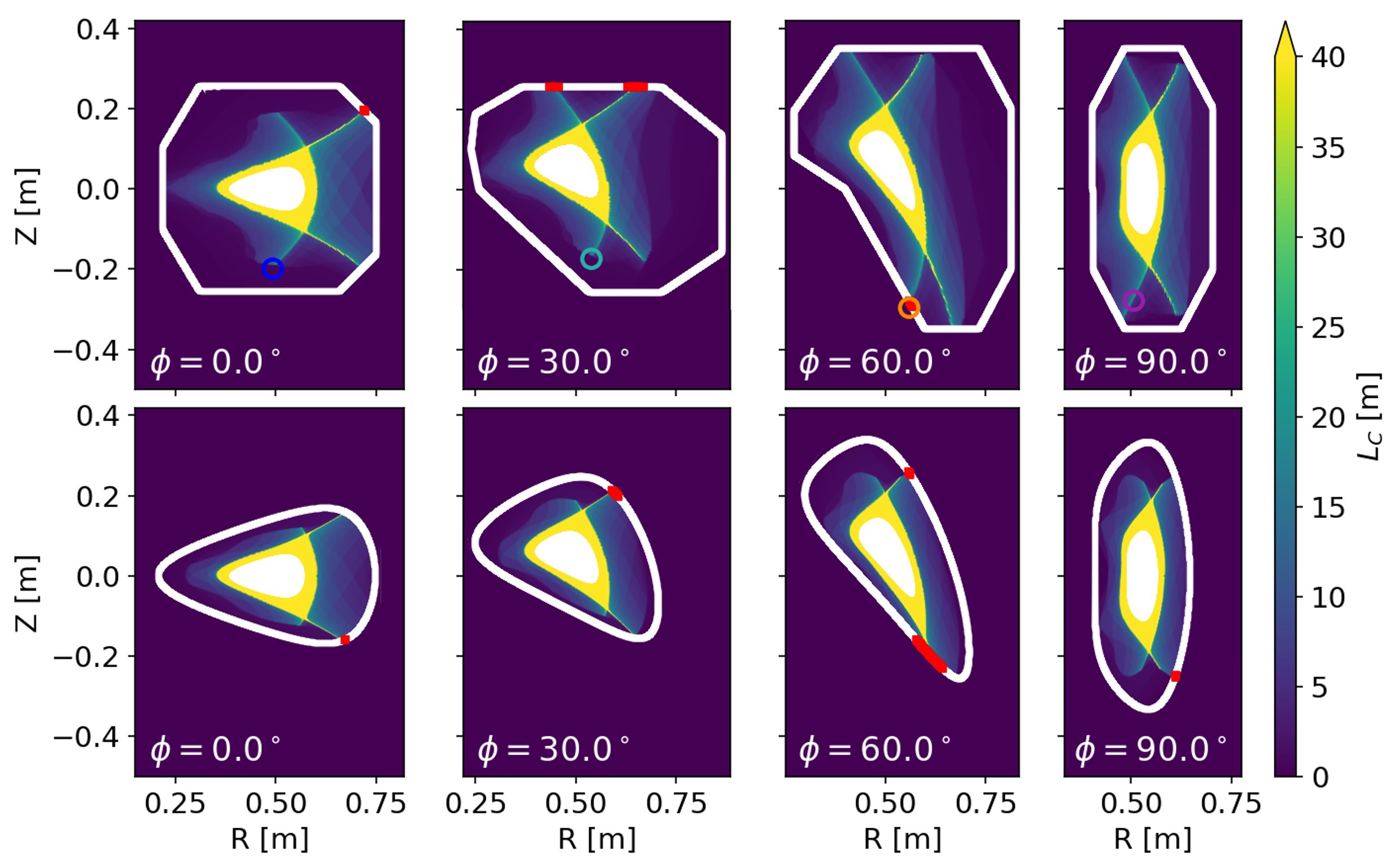}
    \caption{Comparison of magnetic connection lengths and vacuum vessel geometries. Poloidal cross-sections showing the connection length ($L_{C}$) of the medium $\iota$ configuration calculated via EMC3-Lite across four toroidal angles ($\phi = 0.0^\circ, 30.0^\circ, 60.0^\circ, 90.0^\circ$). The region within the Last Closed Flux Surface (LCFS) is omitted (white central region) due to connection lengths exceeding the display scale. The color scale indicates $L_{C}$ ranging from $0.1\,\unit{\meter}$ to above $40\,\unit{\meter}$ in the region enclosed by the X-point manifolds and the divertor legs. Regions along the vessel where the simulated heat flux exceeds $10\,\mathrm{kW/m^2}$ are marked in red. The colored circles in the top row represent the fieldline used to visulaize the concept of shadowing in \Cref{fig:shadowing}.
    }
     \label{fig:lcon}
\end{figure}

\begin{figure}
    \centering
    \includegraphics[width=\textwidth]{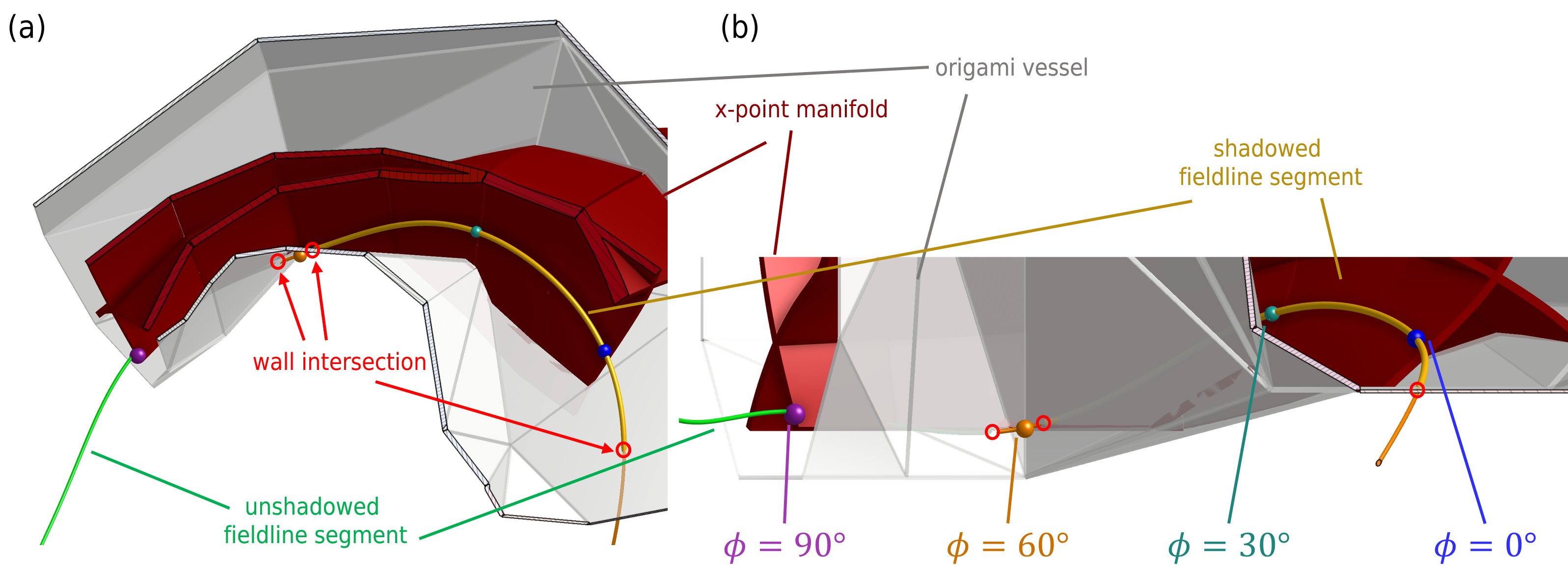}
    \caption{Visualisation of magnetic field line shadowing within the origami vessel, which is cut open at a constant $z$ plane. (a) Isometric view showing the intersection of a single magnetic field line with the vessel wall. (b) Side view displaying the field line at various toroidal angles $\phi$. Shadowing occurs because the field line, lying on the x-point manifold (red), intersects the wall more than once as it moves away from the X-point. These wall intersections (marked with red circles) divide the field line into an unshadowed segment (green, e.g., at $\phi=90^\circ$) that approaches the X-point with a long connection length, and a shadowed segment (yellow, e.g., at $\phi=0^\circ$) which has a very short connection length ($L_c = 36\,$cm) due to vessel intersections%
    }
    \label{fig:shadowing_manifold}
\label{fig:shadowing}
\end{figure}

To illustrate these shadowing effects more clearly, we calculate the target-to-target connection length $L_C$, defined as the total distance that can be traveled along a given field line before being interrupted by a solid component, using EMC3-Lite (see 
\cite{smiet_turnstiles_2025} for details). %
Connection length is often an informative quantity because, for a given magnetic configuration, the largest $L_C$ on the PFCs usually coincides with where the parallel heat fluxes on the PFCs are greatest \citep{effenberg2019investigation, gao_2023, kharwandikar2025power, boeyaert2025towards}. $L_C$ for the two vessels for the medium $\iota$ configuration are shown in \Cref{fig:lcon}. Beyond the confined region (where $L_C$ is infinite), $L_C$ peaks on the four divertor legs, reaching tens of meters (strictly the connection length \textit{on} the manifolds is infinite, because they asymptotically approach the X-point, but $L_C$ exceeds $50\,\unit{\meter}$ only in a tiny region). $L_C$ shows discontinuous changes at certain points along the manifolds as a result of this target shadowing, i.e. dropping sharply from tens of meters to less than a meter. This is best seen on the inner divertor legs at $\phi=0$ in \Cref{fig:lcon}. To illustrate the physical mechanism behind the sharp drop in $L_C$, \Cref{fig:shadowing} isolates a single in-vessel magnetic field line lying on the manifold, which corresponds to the colored circles plotted in the top row of \Cref{fig:lcon}. If the magnetic field and vessel were perfectly axisymmetric, a field line on the manifold would intersect the vessel exactly once and asymptotically approach the X-point with an infinite connection length. However, as shown in \Cref{fig:shadowing}, the 3D geometry of the field and the origami vessel causes the field line to intersect the wall multiple times as it moves away from the X-point. These wall intersections, explicitly marked with red circles, physically divide the field line. The ``unshadowed'' segment (green) successfully approaches the X-point with a long connection length, while the ``shadowed'' segment (yellow) is blocked from accessing the core plasma. Consequently, points on the shadowed segment, such as the blue dot at $\phi=0^\circ$ and the cyan dot at $\phi=30^\circ$ seen in both Figures \ref{fig:lcon} and \ref{fig:shadowing}, exhibit a near-zero ($\approx36\,\mathrm{cm}$) $L_C$ because they terminate at the wall almost immediately when traced. The geometric structure of the divertor legs, combined with connection length information, could be used to design divertor plates which spread the heat toroidally more uniformly. This will be important for reactor-scale devices, in which avoiding PFC thermal overloads can be challenging, and is left as future work.%

The comprehensive analysis of \textbf{design A} confirms that the direct re-optimization procedure yields a configuration that is both physically robust and experimentally flexible. Maintaining high quality quasisymmetry and reliable particle confinement across a wide operational space ensures that STAR\_Lite can move beyond a single equilibrium to explore a range of magnetic field topologies. Most importantly, we have verified that a stable NRD structure exists -- characterized by resilient strike patterns despite the complex 3D constraints of target shadowing, and this demonstrates that STAR\_Lite is well-equipped to meet its primary scientific mission. These results provide the technical assurance necessary to proceed from computational design to physical realization of the STAR\_Lite experiment.

\section{Sensitivity Analysis}
\label{sec:sensitivity}

The third research objective \ref{goal:perturb} is to test the resilience of the X-points and divertor legs to various perturbations of the magnetic field. In the STAR\_Lite experiment, perturbations to the magnetic field have two plausible sources: errors in the fabrication and placement of the coils, and non-zero plasma current. The sensitivity analysis to coil perturbations also informs the engineering design, placing tolerances on construction, and guiding the use of field correction mechanisms. In this section, we numerically evaluate the sensitivity of \textbf{design A} to coil perturbations and plasma current. 

\subsection{Sensitivity to manufacturing errors} 
\label{sec:sensitivity_to_manufacturing}

We now evaluate how manufacturing errors in the coil fabrication process affect quasi-symmetry, rotational transform, and the X-point position. %

Manufactured coils are modeled by adding perturbation $\bm \Xi_i$ to the Cartesian coordinates of the unperturbed coils $\bm \Gamma_i$:
\begin{equation*}
    \tilde{\bm \Gamma}_i(t) = \bm \Gamma_i(t) + \bm \Xi_i(t),
\end{equation*}
where each $\bm \Xi_i(t)$ are different samples from a Gaussian process \citep{wechsung_single-stage_2022,wechsung_stochastic_2022}. 
Note that approach taken here is not equivalent to the one in \cite{glas_global_2022}, where the perturbed device preserves stellarator and discrete rotational symmetry, which is not enforced in the current work. The Gaussian process perturbation model depends on two parameters: $\sigma, \ell$.
The magnitude of the fabrication errors is controlled by the standard deviation parameter, $\sigma$, which we set to $\sigma=1\,\unit{\centi\meter}$ unless otherwise stated. A second parameter, $\ell$, controls the correlation length $\ell$ of the perturbation: larger $\ell$ leads to slower, smoother perturbations. We use a correlation length of $\ell=0.5\,\unit{\meter}$, following previous studies \citep{glas_global_2022, baillod2026update}, as this value produces perturbations that are visually consistent with physical coil deformations. However, since the modular coils are fabricated in-house at Hampton University, we intend to empirically calibrate this parameter in future work. By performing high-resolution 3D scans of the bent steel spines and the finalized windings, we can use the resulting point-cloud data to infer the actual spatial correlation of fabrication errors, ensuring our stochastic models remain grounded in our specific manufacturing process.  
 We can compute the pointwise normal deviation $P(t) = \| (\bm \Xi_i(t) \cdot \mathbf n_i(t)) \mathbf n_i(t) +(\bm  \Xi_i(t) \cdot \mathbf b_i (t))\mathbf b_i(t)\|$, which is Rayleigh $R(\sigma)$ distributed.  This means that the expected normal deviation is $\mathbb{E}[P(t)]=\sigma\sqrt{\pi/2}$ \citep{wechsung_stochastic_2022}, and for $\sigma=1\,\unit{\centi\meter}$, this estimate becomes $\approx 1.25\,\unit{\centi\meter}$. 
  It is worth noting that analyzing the impact of such large deviations is a \textit{worst case} analysis; recently, table-top stellarators with superconducting coils have been designed with much tighter tolerances \citep{baillod2026update,gil_manufacturing_2025}, in alignment with large-scale stellarators like W7-X \citep{rummel2004accuracy}.

\begin{figure}
    \centering
    \includegraphics[width=\linewidth]{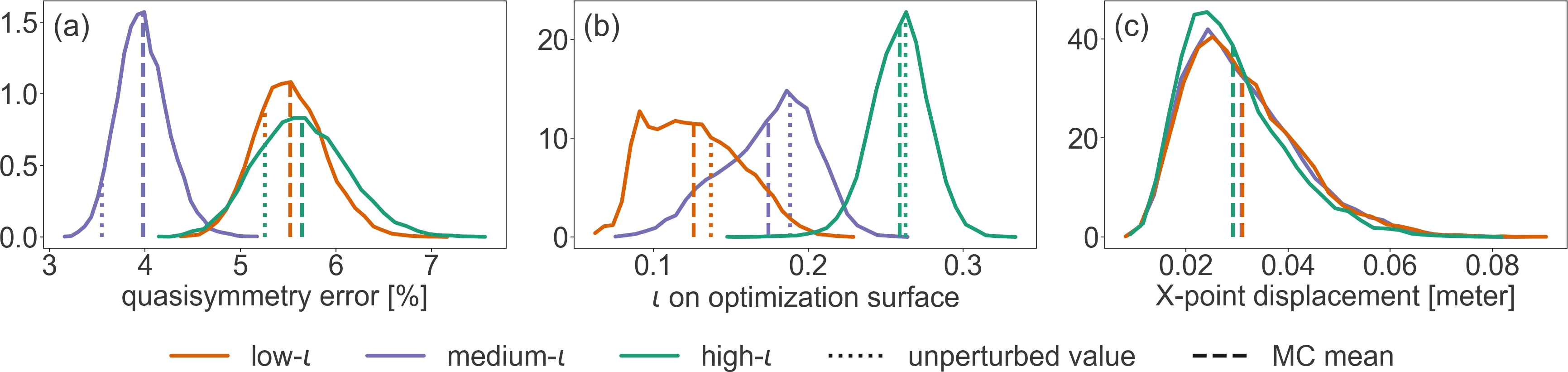}
    \caption{Density of quasisymmetry error (left), rotational transform on the optimization surface (middle), and maximum (over $\phi$) displacement of X-points (right) due to manufacturing errors with ($\sigma=1\,\unit{\centi\meter}, \ell=0.5$).  
    The dotted vertical lines (\dottedline) correspond to the value in the unperturbed \textbf{design A}, while the dashed lines (\dashedline) correspond to the expected value under manufacturing perturbations, computed with over 10,000 Monte Carlo (MC) samples.
    }
    \label{fig:stoch_hist}
\end{figure}

\Cref{fig:stoch_hist} shows the distribution of the quasisymmetry, rotational transform on the optimization surface, and X-point displacement under the Gaussian process manufacturing error model. \Cref{fig:stoch_hist}(a) shows that the expected quasisymmetry error (dashed vertical lines) is not substantially affected by $\mathcal{O}(1\,\text{cm})$  deviations, increasing from the unperturbed design (dotted lines) by less than one percentage point on average. 
Similarly, \Cref{fig:stoch_hist}(c) shows the distribution of the maximum (over $\phi$) X-point displacement, due the manufacturing errors.
We see that when the X-point structure is preserved, the expected displacement is around 3 cm. This encouraging result suggests that achieving  \ref{goal:measurement} is feasible, even under large manufacturing errors. 
While the magnitude of manufacturing errors greatly exceeds the stringent tolerances of large scale devices like W7-X \citep{rummel2004accuracy}, our analysis highlights the resilience of \textbf{design A}. This level of tolerance is particularly advantageous for university-scale prototyping, where it enables successful operation despite the practical constraints of rapid, cost-effective in-house fabrication. 
\Cref{fig:stoch_hist}(b) shows that coil perturbations cause substantial variability of the rotational transform on the optimization surface. Therefore, we expect that precisely obtaining the desired rotational transform values will require implementing a correction mechanism.

For completeness, we have also examined the impact of smaller errors, with $\sigma=1\,\unit{\milli\meter}$ (approximately $1.25\,\unit{\milli\meter}$ expected pointwise deviations), but do not show the associated histograms. The quasisymmetry error degradation, standard deviation of the edge rotational transform, and X-point deviation behave commensurately to the reduction of $\sigma$, decreasing by an order of magnitude.

It is difficult to determine an accurate probability with which nested surfaces and the X-point structure are preserved in the perturbed devices, but a useful proxy for this is the failure rate of our solvers that compute the X-points and Boozer surface.
The Boozer surface solver is robust and can converge even if nested surfaces do not exist in the perturbed magnetic field, though the computed solution will be unphysical.  The X-point solver relies on Newton's method, which should not converge if the NRD structure does not exist or if an unsuitable initial guess is provided to the algorithm.
With these caveats about the solvers in mind, we estimate that the probability of failure of the final build is less than 5\% based on samples drawn from a range of manufacturing errors. When the X-points and nested magnetic surfaces exist, and the solvers converge, it is reassuring to find that the hyperbolic fixed points persist under small perturbations as expected from normally hyperbolic invariant manifold theory~\citep{fenichel1971persistence}.

\begin{figure}
    \centering
    \includegraphics[width=\linewidth]{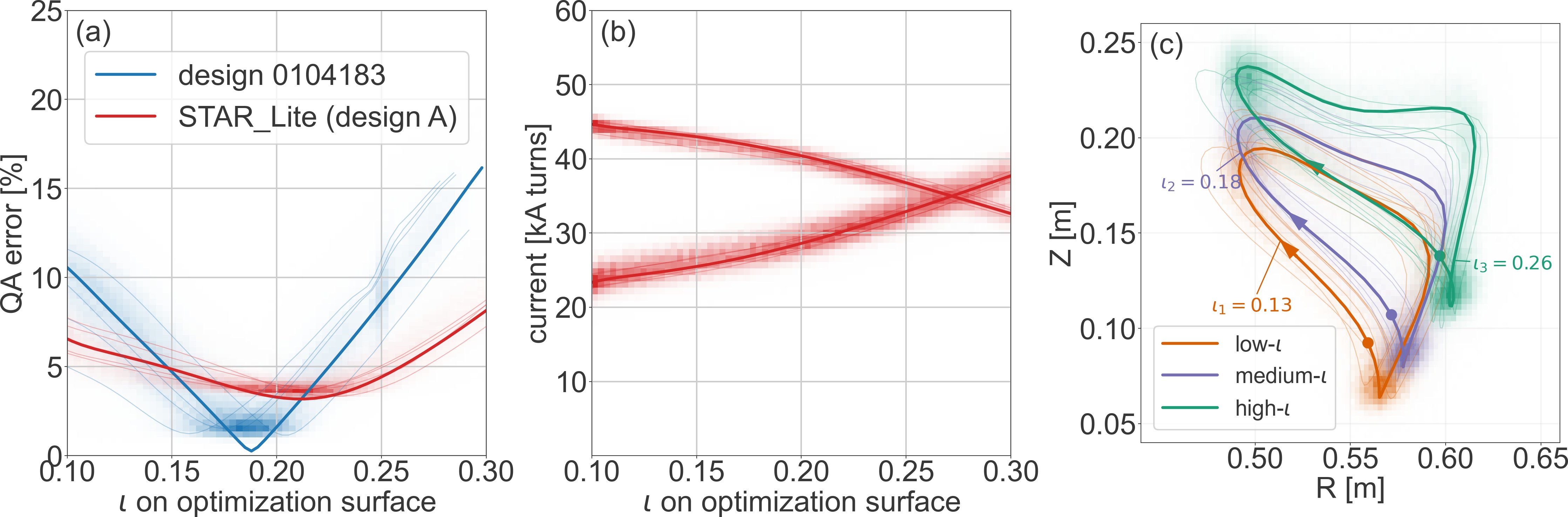}
    \caption{
    A stochastic version of \Cref{fig:qs}. Quasisymmetry error (left), rotational transform (middle), and X-point position (right) for coils with ($\sigma=1\,\unit{\centi\meter}, \ell=0.5$) manufacturing errors. The curves are generated by varying the current ratios between the $L$ and $T$ coils. The thick solid lines correspond to the original unperturbed \textbf{design A}, while the solid thin lines correspond to randomly sampled coil sets. The shaded region represents the distribution of sampled devices, generated using over 4,000 Monte Carlo (MC) samples.}
    \label{fig:stoch}
\end{figure}
We generate a stochastic version of \Cref{fig:qs} in \Cref{fig:stoch}, by randomly sampling over 4,000 coils with manufacturing errors ($\sigma=1\,\unit{\centi\meter}, \ell=0.5\,\unit{\meter}$).  
For each (perturbed) coil set, we compute quasisymmetry error (left), rotational transform (middle), and the X-point curves (right), over a range of current ratios in the $L$ and $T$-coils. The thick solid lines correspond to the original unperturbed \textbf{design A}, while the solid thin lines correspond to randomly sampled devices. The shaded regions emphasize the likelihood of an outcome.
\Cref{fig:stoch} illustrates that although the original current ratios may not give us precisely the desired rotational transform, it appears likely that a different, nearby current ratio exists. Thus, we find that quasisymmetry is relatively insensitive to coil perturbations as large as $1\,\mathrm{cm}$. On the other hand, rotational transform and X-point position can vary significantly, suggesting the need for corrective mechanisms.

\begin{figure}
    \centering
    \includegraphics[width=\linewidth]{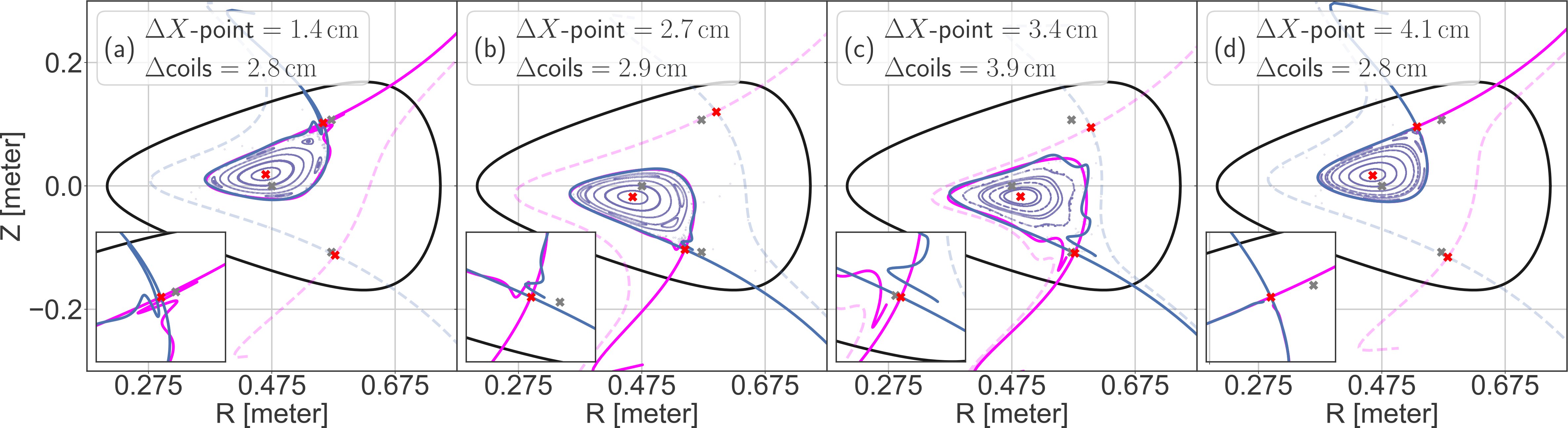}
    \caption{Poincar\'e sections of four randomly sampled devices with different upper X-point deviations in the cylindrical $\phi=0$ plane, $\Delta \text{X-point} = \|\tilde{\bm \Gamma}_{\text{X-point}}(\phi=0)-\bm \Gamma_{\text{X-point}}(\phi=0)\|$, ranging from approximately 1 to $4\,\mathrm{cm}$, labeled at the top left of each panel $\Delta X$-point.  We also quote the maximum normal deviation in the coils, $\max_{t\in [0, 1), i=1,\hdots,6}\| (\bm \Xi_i(t) \cdot \mathbf n_i(t)) \mathbf n_i(t) +(\bm  \Xi_i(t) \cdot \mathbf b_i(t)) \mathbf b_i(t)\|$, where $\mathbf n_i(t), \mathbf b_i(t)$ are the normal and binormal unit vectors on the coil.  The gray crosses correspond to the X-point and axis positions in the unperturbed design A, and the red correspond to the perturbed device. Manifolds of perturbed X-points are also shown as solid and dashed lines.
    The coil currents used here match those used for the medium $\iota$ configuration, and the black cross section corresponds to the extended vacuum vessel.
    }
    \label{fig:poincare}
\end{figure}

In \Cref{fig:stoch_hist}(c), we observe that manufactured devices have a maximum X-point displacement that is on the order of a few centimeters. 
Devices with such magnitude of displacement have strong deviations from stellarator symmetry, as shown in \Cref{fig:poincare}. Magnetic variations can induce chaos into the magnetic field (as shown by the homoclinic tangle in the insets of \Cref{fig:poincare}(a,b,c,d)). They can also perturb the X-points such that the divertor resembles an upper or lower single null divertor, in which the plasma is bounded by the manifolds of one X-point but not the other (i.e. in presence of chaos, the X-point nearer the confined plasma creates a \textit{homoclinic} tangle rather than a \textit{heteroclinic} tangle as in \Cref{fig:manifolds_v2}(d)). 
We also observe in \Cref{fig:poincare}(d) that large coil and X-point deviations do not necessarily result in augmented stochasticity near the top X-point.
\Cref{fig:poincare} suggests that error field correction techniques beyond simply varying the current ratios may have to be implemented to obtain the planned double-null design, if large manufacturing errors are made in coil production.
Possible remediation mechanisms include applying solid body rotations and translations to the manufactured coils \citep{wechsung_stochastic_2022}, and constructing an array of small planar trim coils \citep{Thea_2025}.
The upside of \Cref{fig:poincare} is that STAR\_Lite may be able to experiment on single-null divertor architectures, simply by perturbing the coil configuration. 

\subsection{Sensitivity to plasma pressure and current}
\label{sec:finite_beta}
In this section, we quantify the effect of including plasma on the otherwise vacuum magnetic field. To do so, we introduce pressure and toroidal current profiles $p(s), I(s)$ where $s$ is normalized toroidal flux coordinate spanning from the axis to a boundary surface. We run free boundary VMEC with the $\iota_2$ current configurations using $\texttt{NS}=201$ surfaces, \texttt{FTOL}=$5\times 10^{\sm 14}$, and an \texttt{MGRID} file with $ n_\phi= 512, n_R = 512, n_Z   = 512$ points. The size of the boundary is determined by the toroidal flux, which we set to be the vacuum toroidal flux through the aspect ratio $6.6$ surface used in optimization. We compare the quasisymmetry error, $\iota$, and $X$-point position to that of the vacuum configuration. We expect the vacuum field to be a good approximation to the low density and temperature plasmas in STAR\_Lite.

\begin{table}
  \begin{center}
\def~{\hphantom{0}}
  \begin{tabular}{lcccc}
      Density  & Ion Temp.   &  Electron Temp. & Plasma $\beta$ \\[3pt]
      $5\times 10^{17}\,\unit{\meter^{-3}}$ & $1\,\unit{\electronvolt}$ & $10\,\unit{\electronvolt}$ & $0.01\%$
  \end{tabular}
  \caption{Plasma parameters at desired experimental operating point.}
  \label{tab:plasma_parameters}
  \end{center}
\end{table}
 
We consider plasmas with densities and temperatures similar to the TJ-K stellarator \citep{niedner_numerical_2003}, see \Cref{tab:plasma_parameters}. Profiles of the form,
\begin{align}
    p(s) = p_0(1 - s),
    \label{eq:pressure_profile}
    \\
    I(s) = I_0(s - s^2),
    \label{eq:current_profile}
\end{align}
are valid for small pressures and currents. A current profile that decays to zero at the boundary ensures that the fields generated by the plasma currents will have minimal effect on divertor topology. Assuming equal density of ions and electrons, the pressure is $p_0 = n(T_i + T_e) = 0.88$\,Pa, which leads to $\beta \approx 0.01\%$ for \textbf{design A}. 

We can determine a range of valid plasma currents, $I_0$, through a scaling argument. The perpendicular component of the current density has magnitude $J_\perp  \approx \|\nabla p\| / \|\Bb\| \approx p_0 \|\nabla s\| / B_0 \approx 2p_0 / (aB_0)$. The parallel current on axis satisfies $\pi a^2J_\parallel \approx I_0$.
Generally, we expect $J_\parallel \ll J_\perp$ in stellarators; at the given densities and temperatures, this gives the upper bound, $I_0 \ll 2\pi a p_0 / B_0 = 4.74\,\unit{\ampere}$, which is four orders of magnitude below the coil currents.

The free boundary analysis reveals that quasisymmetry and rotational transform are minimally affected by an increase in pressure and current. In the expected operational conditions ($\beta=0.01\%$, $I_0 = 4\,\unit{\ampere}$), the quasisymmetry error changes by less than $0.01$ percentage points, and the rotational transform changes by roughly $0.001$. Similarly, the maximum (over $\phi$) translation in the $X$-point, due to non-zero pressure in current, is $0.19\,\unit{\milli\meter}$. This change will hardly be noticeable in experiment. In a theoretical, ``high-current, high-pressure'' regime with ten times large plasma pressure and current ($\beta=0.1\%, I_0 = 40\,\unit{\ampere}$), the effect of pressure and current become more noticeable. The upper X-point translates up to $1.85\,\unit{\milli\meter}$, and the rotational transform on the boundary surface shifts by roughly $0.01$. Importantly, even in this extreme case, the  X-point still exists, and could be controlled by changes in the coil currents.

\section{Summary and Outlook}
\label{sec:outlook}

The design and analysis of \textbf{design A}, presented in this work, establishes the foundation for the STAR\_Lite experiment at Hampton University. By prioritizing manufacturing simplicity through the spine-based winding technique, and experimental flexibility through independent coil current control, STAR\_Lite is positioned to validate the NRD concept, experimentally. The realization of this facility will proceed in a phased approach, moving from component fabrication to open-air magnetic validation, and finally to vacuum vessel integration and plasma operations.

The immediate next step is the construction and validation of the design A modular coil set. 
Once the coils are assembled, Phase I of operations will focus on high-fidelity magnetic field mapping, in the absence of a vacuum vessel, using a translatable Hall probe array. This campaign is critical to experimentally verify the magnetic field topology and the existence of the non-resonant divertor structure. 
Mapping the magnetic field structure will identify the effect of manufacturing errors, modeled in \Cref{sec:sensitivity}, and will guide the implementation of field-corrective mechanisms.

Following magnetic validation, the subsequent phase focuses on the detailed design and fabrication of the vacuum vessel. As discussed in \Cref{sec:physics_analysis}, the choice between the ``extended'' and ``origami'' vessel or some compromise of those two extremes involves a trade-off between manufacturing feasibility and edge-physics fidelity. The results from the Hall probe mapping will inform this design process, ensuring the vessel wall conforms to the true magnetic configuration and accommodates the target divertor footprints. Phase II of operations will involve the installation of the vacuum vessel and the $2.45\,\unit{\giga\hertz}$ magnetron heating system. The primary scientific objective of this campaign is to characterize the heat flux footprints on the divertor targets and compare them with the EMC3-Lite predictions (\Cref{fig:heatflux}). %

Looking beyond \textbf{design A}, STAR\_Lite is envisioned not as a static experiment, but as a rapid prototyping facility. The moderate scale ($R_0\approx0.5\,\unit{\meter}$), accessible coil design and simple manufacturing process allow for the future testing of alternative magnetic geometries. By enabling the iterative testing of optimized coilconfigurations, STAR\_Lite aims to de-risk the engineering of stellarators with a variety of divertor topologies and contribute to the development of robust exhaust solutions for future stellarator power plants.

\section{Code availability}
\textbf{Design A} is designed and analyzed using SIMSOPT \citep{landreman_simsopt_2021}. The design and analysis scripts are available on Zenodo \url{https://doi.org/10.5281/zenodo.19056413}

\section{Acknowledgements}
We gratefully acknowledge support from the Simons Foundation (1167550, CL), DOE-FAIR (DE-SC0024443), and DOE-RENEW (DE-SC0025698).
This work has partially funded by the European Union via the Euratom Research and Training Programme (Grant Agreement No 101052200 — EUROfusion). Views and opinions expressed are however those of the author(s) only and do not necessarily reflect those of the European Union or the European Commission. Neither the European Union nor the European Commission can be held responsible for them. The authors acknowledge the use of various AI systems for assistance with manuscript formulation and plotting code. All AI-assisted content was reviewed, edited, and verified by the authors, who take full responsibility for the integrity of the final work.
We thank Chris Smiet, Alkesh Punjabi, and Allen Boozer for the useful discussions.

\bibliography{references}  %

\bibliographystyle{jpp}

\newpage
\appendix

\section{Brief literature review of non-resonant divertors}\label{app:nrd_literature}
A common feature of NRDs in literature is resilient strike locations on plasma-facing components. The concept of resilience was first introduced in early designs of Wendelstein 7-X \cite{strumberger1992magnetic, strumberger1992topology, nuhrenberg1992development}, where it was observed that the plasma boundary showed regions of large curvature which form sharp helical ridges. It was found that field lines outside the plasma boundary increased their radial separation in these high-curvature regions, independent of specific island chains in the edge. A ``helical trough'' divertor was proposed, with divertor plates aligning with sharp edges of the plasma boundary. Similar observations were made in the divertor modeling of NCSX \citep{mioduszewski2007power}, the ARIES-CS reactor concept \citep{mau2008divertor}. However, these examples pre-date the term ``non-resonant divertor'', which first appeared in the literature in 2015 \citep{boozer_2015} and was then applied to NCSX and the ARIES-CS. Since 2015, the term NRD has also been given to HSX \citep{Bader_2017}, CTH \citep{bader2018minimum, Garcia2023CTH} and the WISTELL-A stellarator design \citep{bader_advancing_2020}. Simulations of these revealed helical footprints which show little change with respect to changes in the plasma equilibrium changes (for example, the coil currents changes in HSX \cite{Garcia2025HSX} or plasma current changes in CTH \citep{Garcia2023CTH}).

A mathematical approach to study NRDs was proposed \cite{boozer_2015} by constructing a Hamiltonian system analogous to a stellarator magnetic field, extensively studied \citep{Boozer_2018, Punjabi_2020, punjabi2022magnetic}. In the Hamiltonian model, a similar resilience is found by following trajectories of the Hamiltonian system initialized a small distance outside the LCFS (or initialized within the LCFS and allowed to slowly move radially outwards as they are traced). Such trajectories strike the prescribed vessel wall in a small number of distinct, narrow toroidal bands, which are largely independent of the vessel wall.

Understanding how the magnetic field gives rise to these resilient structures has received considerable attention. One explanation is X-points (periodic magnetic field lines which are hyperbolic i.e. exponentially attract/repel nearby field lines) and/or ``sharp ridges'' which resemble X-points. For HSX, W7-X, NCSX and ARIES-CS it was noticed that the strike patterns tend to align with regions of sharp curvature on the LCFS i.e. where the LCFS contains helical ridges. The helical ridges have been compared with the properties of X-points \citep{boozer_2015}, but it was emphasized that these ridges are discontinuous, and have a rotational transform which is greater than the field lines on the surface (the ridges make approximately half a poloidal rotation per field period). Therefore, these sharp edges cannot strictly be X-points (they are not periodic magnetic field lines). %
A study of \citep{davies_topology_2025} examined the NRD Hamiltonian system and concluded that the diverting structures in this model are a consequence of (actual) X-points, which have a rotational transform of zero. This situation is somewhat similar to the poloidal divertors of tokamaks \citep{stangeby2000tutorial}. 

Another feature often associated with NRDs is the presence of magnetic chaos, that is, regions where magnetic field lines do not lie on surfaces but ergodically sample a volume. Magnetic chaos has been reported and studied to some extent in all of the previous examples. In particular, in the NRD Hamiltonian system, field lines are found to be collimated by partial transport barriers (sometimes referred to as ``cantori'') into escaping bundles known as magnetic turnstiles~\citep{Boozer_2018, Punjabi_2020}. %
The role of magnetic chaos in NRDs and its implications for stellarator divertor performance more generally has also motivated recent ``NRD'' experiments in W7-X, in which a spatially large chaotic region exists in the plasma edge \citep{boeyaert2025analysis}.

\end{document}